\documentclass[prd,twocolumn,nofootinbib,notitlepage]{revtex4-1}
\usepackage[utf8]{inputenc}
\usepackage{cancel}

\usepackage{hyperref}
\usepackage{latexsym}
\usepackage{amsmath}
\usepackage{amssymb}
\usepackage{bbm}

\usepackage{ulem}
\usepackage{pdfsync}
\usepackage{epsfig}
\usepackage{epstopdf}
\usepackage{subfigure}
\usepackage{color}
\usepackage{comment}
\usepackage{slashed}

\def\beq{\begin{equation}}
\def\eeq{\end{equation}}
\def\baq{\begin{eqnarray}}
\def\eaq{\end{eqnarray}}

\newcommand{\be}{\begin{equation}} 
\newcommand{\ee}{\end{equation}}
\newcommand{\bea}{\begin{eqnarray}} 
\newcommand{\eea}{\end{eqnarray}}
\newcommand{\nn}{\nonumber}
\newcommand{\bmp}{\noindent\begin{minipage}{16cm}}
\newcommand{\emp}{\end{minipage}\vskip 7mm} 
\def\lsim{\mathrel{\raise.3ex\hbox{$<$\kern-.75em\lower1ex\hbox{$\sim$}}}}
\def\gsim{\mathrel{\raise.3ex\hbox{$>$\kern-.75em\lower1ex\hbox{$\sim$}}}}

\newcommand{\intron}[1]{}

\def\MP{M_{\rm P}}

\begin{document}

\title{New universal attractor in nonmininally coupled gravity: Linear inflation}

\author{Antonio Racioppi}
\email{antonio.racioppi@kbfi.ee}
\affiliation{National Institute of Chemical Physics and Biophysics, \\
						R\"avala 10, 10143 Tallinn, Estonia}

\begin{abstract}
Once quantum corrections are taken into account, the strong coupling limit of the  $\xi$-attractor models (in {\it metric} gravity) might depart from the usual Starobinsky solution and move into linear inflation. Furthermore, it is well known that the {\it metric} and {\it Palatini} formulations of gravity lead to different inflationary predictions in presence of non-minimally couplings between gravity and the inflaton. In this letter we show that for a certain class of non-minimally coupled models, loop corrections will lead to a linear inflation attractor regardless of the adopted gravity formulation.
\end{abstract}

%
\maketitle

%
\section{Introduction}
According to the theory of cosmic inflation~\cite{Starobinsky:1980te,Guth:1980zm,Linde:1981mu,Albrecht:1982wi}, our Universe underwent a period of exponential expansion during the initial instants of its life. In addition to offering a solution to issues like the flatness and horizon problems of the Universe, inflation also has the merit of providing a way to generate primordial inhomogeneities, whose power spectrum is currently being tested in several experiments~\cite{Ade:2015tva,Ade:2015xua,Ade:2015lrj,Array:2015xqh}. In particular, the latest data from the BICEP2/Keck Collaboration~\cite{Array:2015xqh} cast strong constraints on the tensor-to-scalar ratio $r$, a quantity related to the amplitude of primordial gravitational waves and to the scale of inflation.  As a consequence, the predictions of the linear inflation model for $r$ as a function of the scalar spectral index $n_s$ now lie on the very edge of the $2 \sigma$ boundary constraint, leaving linear inflation as the first model to be either confirmed or ruled out by the upcoming data release.
Linear inflation can be generated via several mechanisms like hilltop inflation \cite{Boubekeur:2005zm}, axion monodromy \cite{McAllister:2008hb}, fermion condensates \cite{Iso:2014gka} and non-minimally coupled to gravity models \cite{Kannike:2015kda,Rinaldi:2015yoa,Barrie:2016rnv,Artymowski:2016dlz,Racioppi:2017spw,Karam:2017rpw}.
In this paper we are going to study models of inflation with a non-minimal coupling to gravity of the type $\xi \phi^2 R$, where $\phi$ is the inflaton field, $R$ the Ricci scalar and $\xi$ a coupling constant. Similar models have been studied in a large number of works over the past decades (in e.g.\cite{Futamase:1987ua,Salopek:1988qh,Fakir:1990eg,Amendola:1990nn,Kaiser:1994vs,Bezrukov:2007ep,Bauer:2008zj,Park:2008hz,Linde:2011nh,Kaiser:2013sna,Kallosh:2013maa,Kallosh:2013daa,Kallosh:2013tua,Kallosh:2013tua,Rubio:2014wta,Csaki:2014bua,Galante:2014ifa,Chiba:2014sva,Boubekeur:2015xza,Pieroni:2015cma,Jarv:2016sow,Salvio:2017xul,Rasanen:2017ivk,Tenkanen:2017jih,Racioppi:2017spw,Karam:2017rpw}). 
These models are particularly interesting, since non-minimal couplings should be seen as a generic ingredient of consistent model frameworks, arising from quantum corrections in a curved space-time \cite{Birrell:1982ix}. In particular, this is the case for the scenario where the Standard Model Higgs boson is the inflaton field \cite{Bezrukov:2007ep}.
Comparisons of non-minimally coupled chaotic models of inflation were performed in e.g.\ \cite{Linde:2011nh,Kaiser:2013sna,Kallosh:2013maa,Kallosh:2013daa,Kallosh:2013tua,Galante:2014ifa,Jarv:2016sow}. In Refs. \cite{Kaiser:2013sna,Kallosh:2013tua}, it was discovered that for large values of the non-minimal coupling, all models, independently of the original scalar potential, asymptote to a universal attractor: the Starobinsky model \cite{Starobinsky:1980te}. 

The introduction of non-minimal couplings to gravity requires a discussion of what are the gravitational degrees of freedom. In the common metric formulation of gravity the independent variables are the metric and its first derivatives, while in the Palatini formulation the independent variables are the metric and the connection. Starting from the Einstein-Hilbert Lagrangian, the two formalisms present the same equations of motion and therefore describe equivalent physical theories. However, in presence of non-minimal couplings between gravity and matter, such equivalence is lost and the two formulations describe different theories of gravity \cite{Bauer:2008zj} and lead to different phenomenological predictions, as recently investigated in e.g. \cite{Tamanini:2010uq,Bauer:2010jg,Rasanen:2017ivk,Tenkanen:2017jih,Racioppi:2017spw,Markkanen:2017tun,Jarv:2017azx}. 
In particular, in \cite{Jarv:2017azx} it has been shown that the attractor behaviour of the so-called $\xi$-attractor models \cite{Kallosh:2013tua} is lost in the Palatini formulation.  It is important to stress that in \cite{Kallosh:2013tua,Jarv:2017azx} the role of quantum corrections is implicitly assumed to be subdominant. On the other side, it has been demonstrated that quantum corrections to inflationary potentials may play a relevant role \cite{Kannike:2014mia,Marzola:2015xbh,Marzola:2016xgb,Dimopoulos:2017xox}, dynamically generating the Planck scale \cite{Kannike:2015apa,Kannike:2015kda}, predicting super-heavy dark matter \cite{Farzinnia:2015fka,Kannike:2016jfs} and leading to linear inflation predictions when a non-minimal coupling to gravity is added \cite{Kannike:2015kda,Rinaldi:2015yoa,Barrie:2016rnv,Artymowski:2016dlz,Racioppi:2017spw}. 

The purpose of this letter is to present a class of non-minimally coupled models where, because of the aforementioned quantum corrections, the strong coupling limit leads {\it universally}, i.e. independently of the gravity formulation, to a linear inflation attractor.


\section{Non-minimally coupled models}
\label{inflation}
Consider the following action of the scalar-tensor theory with the flat FRW metric tensor
\begin{equation}
S = \!\! \int \!\! d^4x \sqrt{-g}\left(-\frac{M_P^2}{2}f(\phi)R(\Gamma) + \frac{(\partial \phi)^2}{2}  - V(\phi) - \Lambda \right) ,
\label{eq:JframeL}
\end{equation}
where $M_P$ is the reduced Planck mass, $R$ is the Ricci scalar constructed from a connection $\Gamma$ and $V(\phi)$ is the inflaton scalar potential. 
The cosmological constant $\Lambda$ is adjusted so that at the minimum the potential value is zero, i.e.
 \begin{equation}
  V_\text{eff}(v_\phi)= V (v_\phi) + \Lambda = 0 \label{eq:Vmin} \, .
 \end{equation}
Our focus is on quartic inflaton potentials\footnote{A more general discussion considering different types of potentials, non-minimal couplings and also non-minimal kinetic terms \cite{Koivisto:2005yc} of the type $K(\phi) \partial_\mu \phi \partial^\mu \phi$ is beyond the purpose of the present letter and it will be presented in a future article.} where loop corrections (coming from other particles like reheating products, UV completions, etc.)
are relevant. The 1-loop effective potential\footnote{It has been proven that the cosmological perturbations are invariant under
a change of frame (see for instance \cite{Prokopec:2013zya,Jarv:2016sow}). On the other hand, the quantum equivalence of the Einstein and Jordan
frames is still an unsolved issue. In the present letter we adopt the following computational strategy: we assume that the effective potential in the Jordan frame is given by eqs.  (\ref{eq:Veff}) and (\ref{eq:lrun}). Once we have the final expression of the 1-loop Jordan frame scalar potential, we move to the Einstein frame, where the calculation of the slow-roll parameters is easier. Given a scalar potential in the Jordan frame, the cosmological perturbations are then independent, in the slow-roll approximation, from the choice of the frame in which we perform the inflationary calculations \cite{Prokopec:2013zya,Jarv:2016sow}. For further readings on frame equivalence and/or loop corrections in scalar-tensor theories see Refs. \cite{Jarv:2014hma,Kuusk:2015dda,Kuusk:2016rso,Flanagan:2004bz,Catena:2006bd,Barvinsky:2008ia,DeSimone:2008ei,
Barvinsky:2009fy,Steinwachs:2011zs,Chiba:2013mha,George:2013iia,Postma:2014vaa,
Kamenshchik:2014waa,George:2015nza,Miao:2015oba,Inagaki:2015fva,Burns:2016ric,
Hamada:2016onh,Fumagalli:2016lls,Fumagalli:2016sof,Bezrukov:2017dyv,Karam:2017zno,Narain:2017mtu,Ruf:2017xon,Markkanen:2017tun,Ferreira:2018itt}.} of the inflaton can be written in a model independent way as:
\begin{equation}
  V(\phi) = \frac{1}{4} \lambda(\phi)\phi^4
  \label{eq:Veff}
\end{equation}
where $\lambda(\phi)$ is a self quartic coupling, whose 
running is described by its beta function
\begin{equation}
\beta_\lambda (\mu) = \frac{d\lambda}{d\log \mu} \, \label{eq:RGElambda},
\end{equation}
where $\mu$ is the renormalization scale. Ignoring all the details of the inflaton interactions or of the theory completion, we can still solve eq. (\ref{eq:RGElambda}) as a Taylor series
\begin{equation}
\lambda(\phi) = \lambda(\mu_0) + \sum_{k=1}^\infty \frac{\beta_k}{k!} \log^k \left(\frac{\phi}{\mu_0}\right) \label{eq:lambdafull}
 \, ,
\end{equation}
where $\lambda(\mu_0)$ is the value of $\lambda$ at a scale $\mu_0$ and the $\beta_k$ parameters represent the $k$-th derivatives of the beta function evaluated at the scale $\mu_0$.
We assume that we obtain a good approximation of the expansion in (\ref{eq:lambdafull}) by keeping only the first order correction
\begin{equation}
  \lambda(\phi) \simeq \lambda(\mu_0) \left[1 + \delta(\mu_0) \ln\left(\frac{\phi}{\mu_0}\right)\right] \, ,
  \label{eq:lrun}
\end{equation}
where the relative loop correction, $\delta=\beta_{\lambda}/\lambda$, is regarded as a free parameter in this model independent approximation. 
For convenience
we fix the reference scale\footnote{The choice is just a convenient parametrization. In the region of validity of the first order approximation (where $\beta_\lambda$ is essentially constant), the result is independent on the choice of $\mu_0$. The parametrization using $\mu_0 =  M_P$,  is related to another one using $\mu_0 = \mu^* \neq M_P$ via the RGE solution
\bea
\lambda(M_P) &=&  \lambda(\mu^*) \left[1 + \delta(\mu^*) \ln\left(\frac{\MP}{\mu^*}\right)\right] \, ,\nn\\
\delta(M_P) &=& \frac{\beta_\lambda (M_P)}{\lambda (M_P)} = \frac{\beta_\lambda (\mu^*)}{\lambda (M_P)},  \nn
\eea
where we used $\beta_\lambda (M_P) \simeq \beta_\lambda (\mu^*)$.
} $\mu_0 = M_P$.
The potential $V(\phi)$ has been projected onto the direction of inflation, i.e. the direction obtained by setting any other scalar field at the minimum of the potential.
In order to avoid cumbersome notation we will henceforth leave the argument ``$(M_P)$'' understood and write it only when explicitly needed.

Let us discuss now the gravitational sector and its non-minimal coupling to the inflaton.
In order to avoid repulsive gravity we require $f(\phi) > 0$. 
This feature is independent on the eventual gravity formulation (metric or Palatini).
In the metric formulation the connection is determined uniquely as a function of the metric tensor, i.e.\ it is the Levi-Civita connection $\bar{\Gamma}=\bar{\Gamma}(g^{\mu\nu})$
\begin{eqnarray}
\label{eq:LC}
\overline{\Gamma}^{\lambda}_{\alpha \beta} = \frac{1}{2} g^{\lambda \rho} \left( \partial_{\alpha} g_{\beta \rho}
+ \partial_{\beta} g_{\rho \alpha} - \partial_{\rho} g_{\alpha \beta}\right) \, .
\end{eqnarray}
On the other side, in the Palatini formalism both $g_{\mu\nu}$ and $\Gamma$ are treated as independent variables, and the only assumption is that the connection is torsion-free, $\Gamma^\lambda_{\alpha\beta}=\Gamma^\lambda_{\beta\alpha}$. Solving the equations of motion leads to \cite{Bauer:2008zj}
\begin{eqnarray}
\Gamma^{\lambda}_{\alpha \beta} = \overline{\Gamma}^{\lambda}_{\alpha \beta}
+ \delta^{\lambda}_{\alpha} \partial_{\beta} \omega(\phi) +
\delta^{\lambda}_{\beta} \partial_{\alpha} \omega(\phi) - g_{\alpha \beta} \partial^{\lambda}  \omega(\phi) \, ,
\label{eq:conn:J}
\end{eqnarray}
where
\begin{eqnarray}
\label{omega}
\omega\left(\phi\right)=\ln\sqrt{f(\phi)} \, .
\end{eqnarray}
Because the connections (\ref{eq:LC}) and (\ref{eq:conn:J}) different, the metric and Palatini formulation provide indeed two different theories of gravity.
Another way of seeing the differences is to study the problem in the Einstein frame by means of the conformal transformation
\begin{eqnarray}
\label{eq:gE}
g_{\mu \nu}^E = f(\phi) \ g_{\mu \nu} \, .
\end{eqnarray}
In the Einstein frame gravity looks the same in both the formalisms (see also eq. (\ref{eq:conn:J})), however the matter sector (in our case $\phi$) behaves differently. Performing the computations \cite{Bauer:2008zj}, the Einstein frame Lagrangian becomes
\begin{equation}
  \sqrt{- g^{E}} \mathcal{L}^{E} = 
  \sqrt{- g^{E}} \Bigg[  - \frac{M_P^2}{2} R  + 
  \frac{(\partial \chi)^{2}}{2} - U(\chi) \Bigg] \, ,
   \label{eq:Einstein:Lagrangian}
\end{equation}
where $\chi$ is canonically normalized scalar field in the Einstein frame, and its scalar potential is
\be
U(\chi) = \frac{V_\text{eff}(\phi(\chi))}{f^{2}(\phi(\chi))} \, .
\label{eq:U}
\ee
In case of the metric formulation, $\chi$ is derived by integrating the following relation
\begin{equation}
\frac{\partial \chi}{\partial \phi} = \sqrt{\frac{3}{2}\left(\frac{M_P}{f}\frac{\partial f}{\partial \phi}\right)^2+\frac{1}{f}} \, ,  
  \label{eq:dphiE}
\end{equation}
where the first term comes from the transformation of the Jordan frame Ricci scalar and the second from the rescaling of the Jordan frame scalar field kinetic term. 
On the other hand, for the Palatini formulation, the field redefinition is induced only by the rescaling of the inflaton kinetic term i.e.
\begin{equation}
\frac{\partial \chi}{\partial \phi} = \sqrt{\frac{1}{f}} \, ,  
  \label{eq:dphiP}
\end{equation}
where there is no contribution from the Jordan frame Ricci scalar. Therefore we can see that the difference between the two formulations in the Einstein frame relies on the different definition of $\chi$ induced by the different non-minimal kinetic term involving $\phi$.

In the following we will focus on two particular types of $f$ functions. The first one is the usual Higgs-inflation \cite{Bezrukov:2007ep,Bezrukov:2017dyv} non-minimal coupling\footnote{The non-minimal coupling $\xi$ is usually subject to loop corrections as well, parametrized by a beta-function behaving like
\begin{equation}
  16 \pi^2 \beta_{\xi} \approx \xi \sum_k \lambda_k  \, \nn ,
\end{equation}
where $\sum_k \lambda_k$ includes the contribution of other couplings from the scalar sector, for example the ones generating the running of $\lambda$. In order to ignore such quantum corrections, the condition $\beta_{\xi} \ll \xi $ must be satisfied. This has been explicitly realized in \cite{Kannike:2015apa,Kannike:2015kda,Marzola:2015xbh,Marzola:2016xgb}.
Because of the constraint on the amplitude of scalar perturbations (\ref{cobe}),
perturbativity of the theory and the $16 \pi^2$ suppression factor, we assume that such condition is valid also in our model independent construction.}
\begin{equation}
f=1 + \xi \frac{\phi^2}{M_P^2} \, ,  
  \label{eq:f:H}
\end{equation}
where we relaxed the condition that the inflaton is the Higgs boson and allowed the possibility that inflation is driven by another scalar beyond the Standard Model particle content.
The second one
\begin{equation}
f= \xi \frac{\phi^2}{M_P^2} \, ,  
  \label{eq:f:ind}
\end{equation} 
is the extension of the previous non-minimal coupling to the induced gravity \cite{PhysRevLett.42.417,SMOLIN1979253,Venturi:1981,SPOKOINY198439,CervantesCota:1994zf,CervantesCota:1995tz} scenario.

\subsection{Higgs-inflation-like models} \label{subsec:Higgs}

In this subsection we study the phenomenological implications of the non-minimal coupling in eq. (\ref{eq:f:H}). Since the detailed discussion of the reheating mechanism is beyond the purpose of the present letter, we do not need to specify the exact shape of the potential around its minimum. In this case it is sufficient to assume that the loop correction in eq. (\ref{eq:lrun}) does not induce a relevant VEV for the inflaton (i.e. $v_\phi \simeq 0$ and therefore $\Lambda \simeq 0$) and that during inflation the potential is well described by eqs. (\ref{eq:Veff}) and (\ref{eq:lrun}). The corresponding Einstein frame scalar potential is given by
\be
 U(\chi)  =     \frac{\lambda \, M_P^4 \, \phi(\chi)^4}{4\left[ M_P^2 + \xi \phi(\chi)^2 \right]^2} \left[1 + \delta \ln\left(\frac{\phi(\chi)}{\MP}\right)\right] ,
\label{eq:U:H}
\ee
where the difference between the metric and the Palatini formulations is given by the different solution of eqs. (\ref{eq:dphiE}) and (\ref{eq:dphiP}). 
However, it can be shown that in the strong coupling limit, $\xi \to +\infty$, the two formulations share a similar approximated solution
\begin{equation}
\chi \simeq \frac{M_P}{q} \left[\frac{1}{\delta }+\ln \left(\frac{\phi }{M_P}\right)\right]
  \label{eq:chi}
\end{equation}
(where we conveniently chose the zero-value of the Einstein frame scalar field) and a similar approximated Einstein frame potential
\be
U(\chi) \simeq     \frac{\lambda \, \delta \, M_P^3 \, q}{4 \, \xi ^2} \chi 
\label{eq:U:H:chi}
\ee 
where $q$ is either
\begin{equation}
q = q_m = \sqrt{\frac{\xi}{1+ 6 \xi}} \, ,
  \label{eq:q:metric}
\end{equation}
for the metric formulation, or 
\begin{equation}
q = q_P = \sqrt{\xi} \, ,
  \label{eq:q:Palatini}
\end{equation}
for the Palatini formulation. We can see that in both cases the attractor solution is linear inflation, with the only difference in the normalization factor $q$. 
For completeness, we anyway perform a full inflationary analysis considering also $\xi$ values other than the strong coupling limit.
Assuming slow-roll, the inflationary dynamics is described by the usual slow-roll parameters and the total number of e-folds during inflation\footnote{The exact number of e-folds depends on the reheating mechanism and it might be used for discriminating between the metric and the Palatini formulations \cite{Racioppi:2017spw}.  Here we concentrate only on the dynamics during inflation, being the reheating analysis beyond the scope of the present letter.}. The slow-roll parameters are defined as
\be
\epsilon \equiv \frac{1}{2}M_{\rm P}^2 \left(\frac{1}{U}\frac{{\rm d}U}{{\rm d}\chi}\right)^2 \,, \quad
\eta \equiv M_{\rm P}^2 \frac{1}{U}\frac{{\rm d}^2U}{{\rm d}\chi^2} \,,
\ee
and the number of e-folds as
\be
N = \frac{1}{M_{\rm P}^2} \int_{\chi_f}^{\chi_i} {\rm d}\chi \, U \left(\frac{{\rm d}U}{{\rm d} \chi}\right)^{-1},
\label{Ndef}
\ee
where the field value at the end of inflation, $\chi_f$, is defined via $\epsilon(\chi_f) = 1$.  
The field value $\chi_i$ at the time a given scale left the horizon is given by the corresponding $N$. 
To reproduce the correct amplitude for the curvature power spectrum, the potential has to satisfy \cite{Lyth:1998xn,Ade:2015xua}
\be
\label{cobe}
\frac{U(\chi_i)}{\epsilon(\chi_i)} = (0.027M_{\rm P})^4 ,
\ee
and the other two main observables, i.e. the spectral index and the tensor-to-scalar ratio are expressed in terms of the slow-roll parameters by
\bea
n_s &\simeq& 1+2\eta-6\epsilon \\ \nonumber
r &\simeq& 16\epsilon ,
\eea
respectively. 
\begin{figure}[t!]
  \begin{center}
\includegraphics[width=0.45 \textwidth]{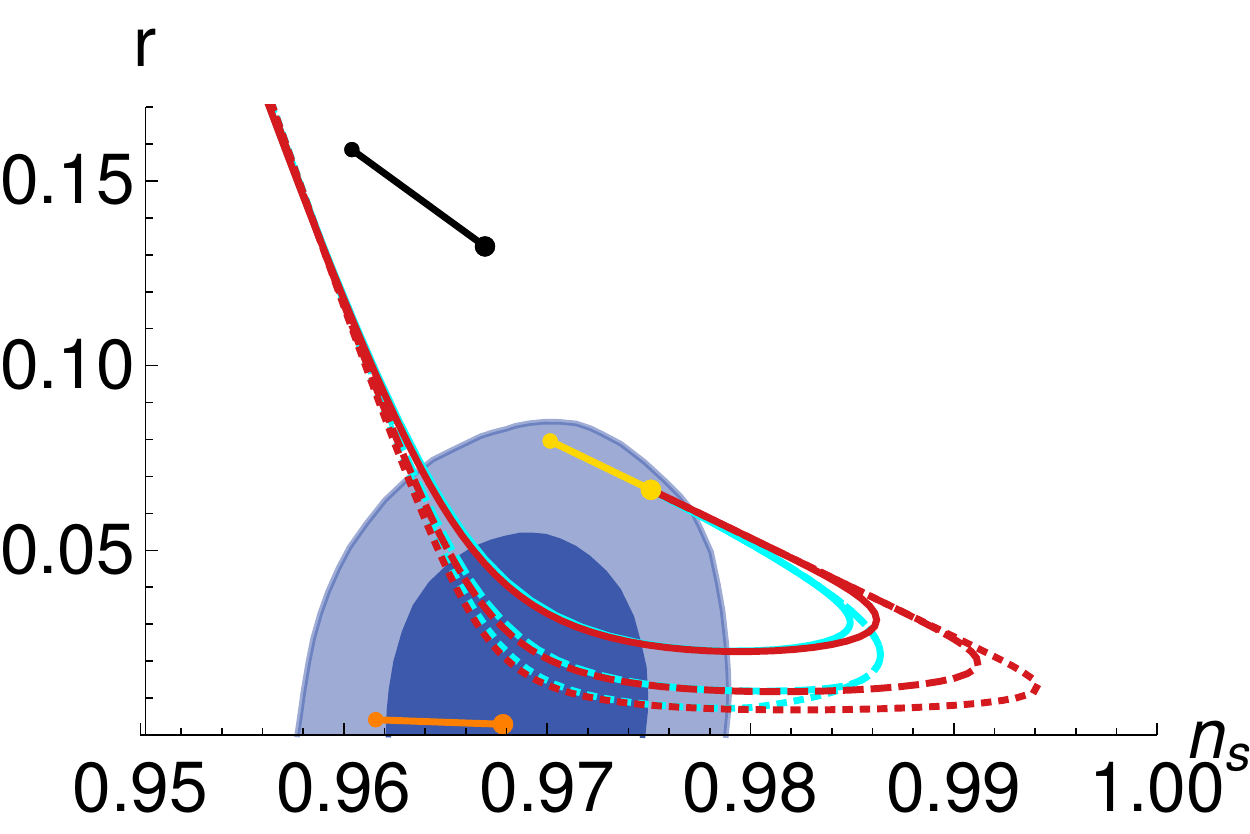}\\
\quad\\
\quad\\
\includegraphics[width=0.45 \textwidth]{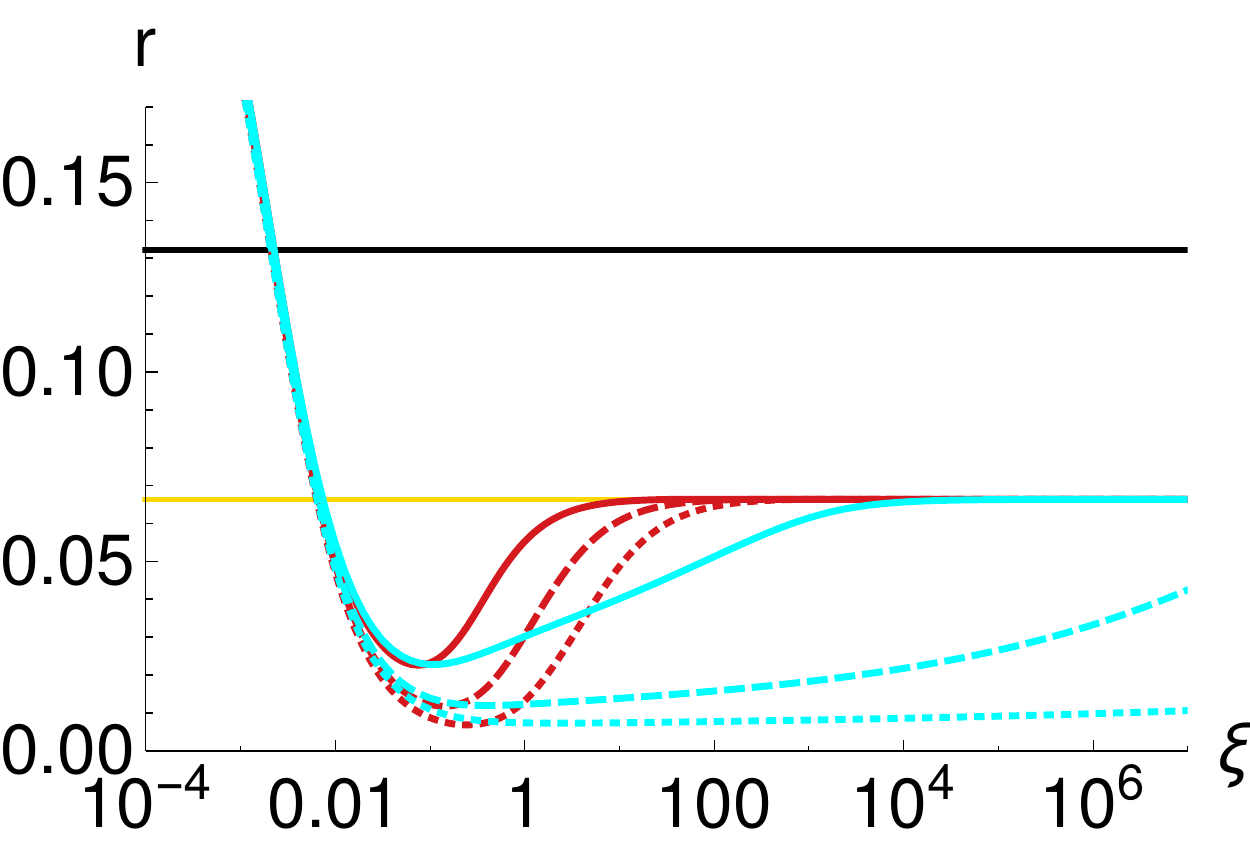}\\
\quad\\
\quad\\
\hspace{-0.3cm}\includegraphics[width=0.45 \textwidth]{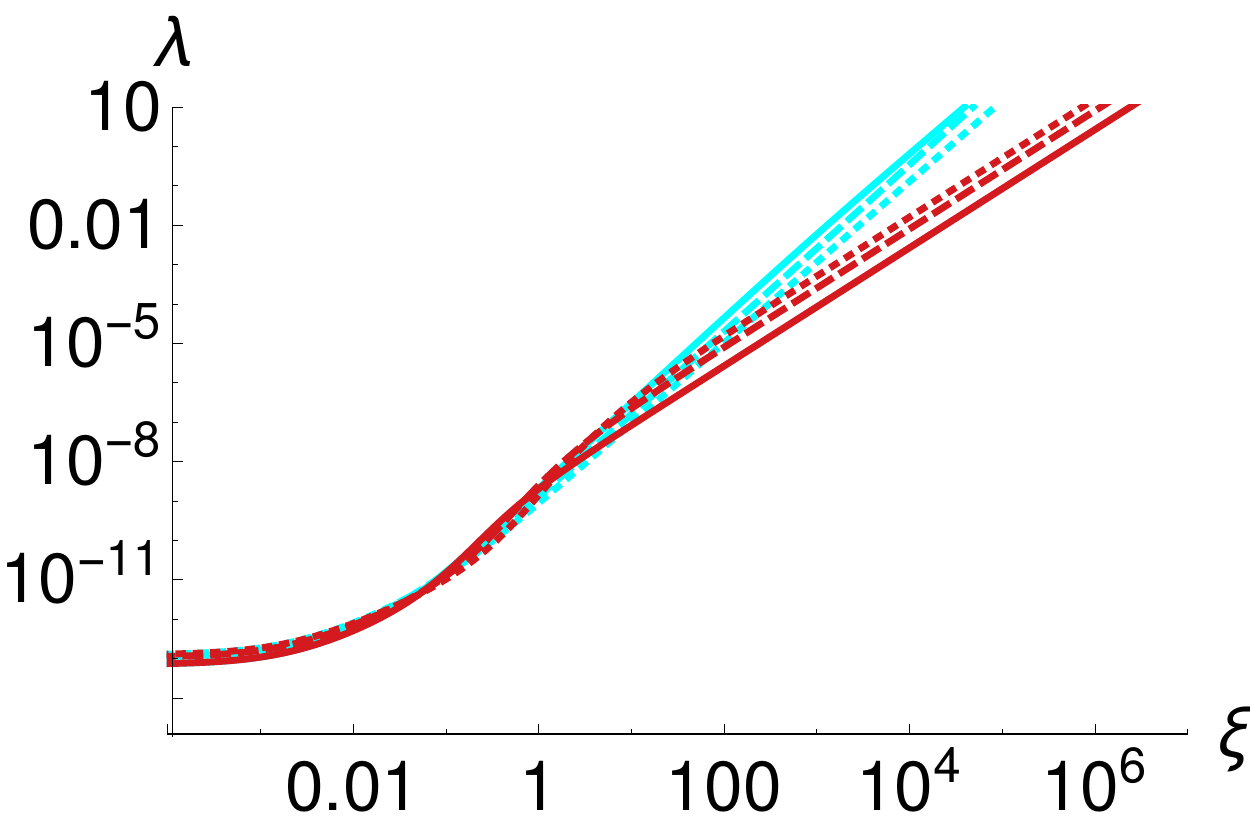}\\
\quad\\
\quad\\
\caption{$r$ vs. $n_s$ (upper panel), $r$ vs. $\xi$ (central panel) and $\lambda$ vs. $\xi$ (lower panel) for $N = 60$ $e$-folds in the metric (cyan) and Palatini formulation (red), with $\delta=5\%$ (dotted), $\delta=10\%$ (dashed) and $\delta=30\%$ (continuous) in the loop corrected Higgs-inflation-like scenario. For reference, we also plot the predictions of quadratic (black), linear (yellow) and Starobinsky (orange) inflation in metric gravity.  The light blue areas present the 1 and 2$\sigma$ constraints from the BICEP2/Keck data \cite{Array:2015xqh}. 
    }
   \label{Fig:r:vs:ns:H}
  \end{center}
\end{figure}
The corresponding results are given in Fig.~\ref{Fig:r:vs:ns:H}, where we depict $r$ vs. $n_s$ (upper panel), $r$ vs. $\xi$ (central panel) and $\lambda$ vs. $\xi$ (lower panel) for $N = 60$ $e$-folds in the metric (cyan) and Palatini formulation (red). Being $\delta$ a relative loop correction we expect it to be smaller\footnote{Such a bound comes from the requirement assumed in the beginning of this subsection, $v_\phi\simeq 0$, which implies  $\delta \ll \frac{4}{2 \ln \xi -1}$ (cf. eqs. (\ref{eq:VEV}) and (\ref{eq:delta})). The region of interest, i.e. the one where we reach the linear attractor, is $\xi \gg 1$, which implies $\delta \ll 1$. Such an upper bound for $\delta$ is also in agreement with the requirement of a theory always perturbative during all the duration of inflation and with the approximation of eq. (\ref{eq:lambdafull}) with eq. (\ref{eq:lrun}).} than 1, at least in the region of validity of eq. (\ref{eq:lrun}). Therefore we decided to plot our results for the following reference values: $\delta=5\%$ (dotted), $\delta=10\%$ (dashed) and $\delta=30\%$ (continuous). For reference, we also plot the predictions of quadratic (black), linear (yellow) and Starobinsky (orange) inflation in metric gravity.  The light blue areas present the 1 and 2$\sigma$ constraints from the BICEP2/Keck data \cite{Array:2015xqh}.
Both formulations share the same behaviour.  First, for small $\xi$ ($\xi \lesssim 0.1$), the predictions are aligned with the strong-coupling limit of the standard (without loop corrections) non-minimal inflation \cite{Kallosh:2013tua}, then, for large $\xi$ values, the loop correction becomes relevant and the results are departing from the Starobinsky attractor and approaching the linear limit. The bigger $\delta$, the sooner the predictions departs from the Starobinsky solution.
We also notice that in the weak ($\xi \ll 1$) and strong ($\xi \gg 1$) coupling limit, which corresponds to the BICEP2/Keck allowed region, the predictions for $r$ vs. $n_s$ of the metric and Palatini formulations essentially overlap. The only way to discriminate between the two formulations would be to take into account the reheating phenomenology and the relation between the exact number of $e$-folds and $\xi$ \cite{Racioppi:2017spw}. To conclude, we also notice that while the Palatini formulation remains always perturbative ($\lambda < 4\pi$) until the linear limit, the metric formulation keeps perturbativity until the linear limit\footnote{This improves the previous analysis of \cite{Marzola:2016xgb,Artymowski:2016dlz} where perturbativity was not kept until the linear inflation limit because of the use of a too small $\delta$.} only for $\delta = 30\%$.



\subsection{Induced gravity models} \label{subsec:induced}
In this subsection we study the phenomenological implications of the non-minimal coupling in eq. (\ref{eq:f:ind}). In this case we assume that the potential is well described by eqs. (\ref{eq:Veff}) and (\ref{eq:lrun}) not only in the inflationary region but also around the minimum of the potential.
Therefore the loop correction in eq. (\ref{eq:lrun}) induces an inflaton VEV
\be
 v_\phi = e^{-\left(\frac{1}{4}+\frac{1}{\delta } \right)} M_P = \frac{M_P}{\sqrt\xi}
\label{eq:VEV}
\ee 
which generates dynamically the Einstein-Hilbert term otherwise missing in eq. (\ref{eq:JframeL}). We can see that in this scenario the relative loop correction $\delta$ and the non-minimal coupling $\xi$ are correlated via eq. (\ref{eq:VEV}). Notice that the requirement $v_\phi^2 >0$ allows also for negative $\delta$. As $\delta=\beta_\lambda/\lambda$, the stability of the potential and therefore consistency of the model is ensured by the constraint $\lambda \delta = \beta_\lambda >0$. We can use the relation (\ref{eq:VEV}) to remove the $\delta$ dependence of the potential, obtaining the following Einstein frame scalar potential
\bea
&&U \!= \! \frac{\lambda  M_P^4}{4 \xi ^2
   \phi ^4} \left[\frac{M_P^4}{\xi ^2 (2 \ln \xi -1)}+
   \left(1+\frac{4 \ln \left(\frac{\phi }{M_P}\right)}{2 \ln \xi -1}\right)\phi ^4 \right] \! , \qquad
\label{eq:U:ind}
\eea
where we used eqs. (\ref{eq:Vmin}) and (\ref{eq:VEV}) to express $\Lambda$ as a function $\lambda$ and $\xi$.
As before, the difference between the metric and the Palatini formulations is given by the different expressions for $\phi(\chi)$. Taking into account that $\delta$ is not any more a free parameter but (see eq. (\ref{eq:VEV}))
\be
\delta=\frac{4}{2 \ln \xi -1} \, ,
\label{eq:delta}
\ee 
it is easy to check that also in this case the field redefinition of eq. (\ref{eq:chi}) and the strong coupling limit ($\xi \to + \infty$) approximation for the potential in eq. (\ref{eq:U:H:chi}) still hold. Therefore, also in this case we generate a linear inflation attractor independently of the adopted gravity formulation. 
For completeness, we perform again a full inflationary analysis considering also $\xi$ values outside the strong coupling limit.
\begin{figure}[t!]
  \begin{center}
\includegraphics[width=0.45 \textwidth]{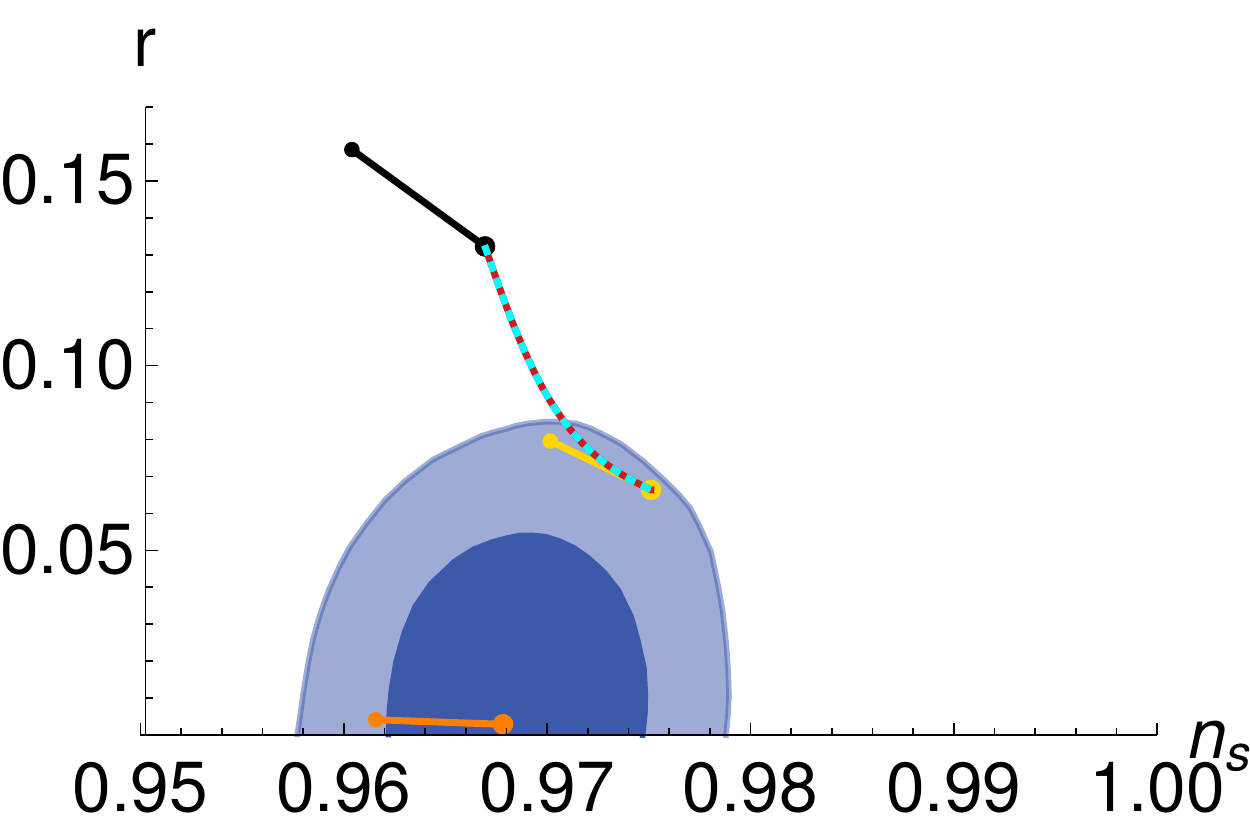}\\
\quad\\
\quad\\
\includegraphics[width=0.45 \textwidth]{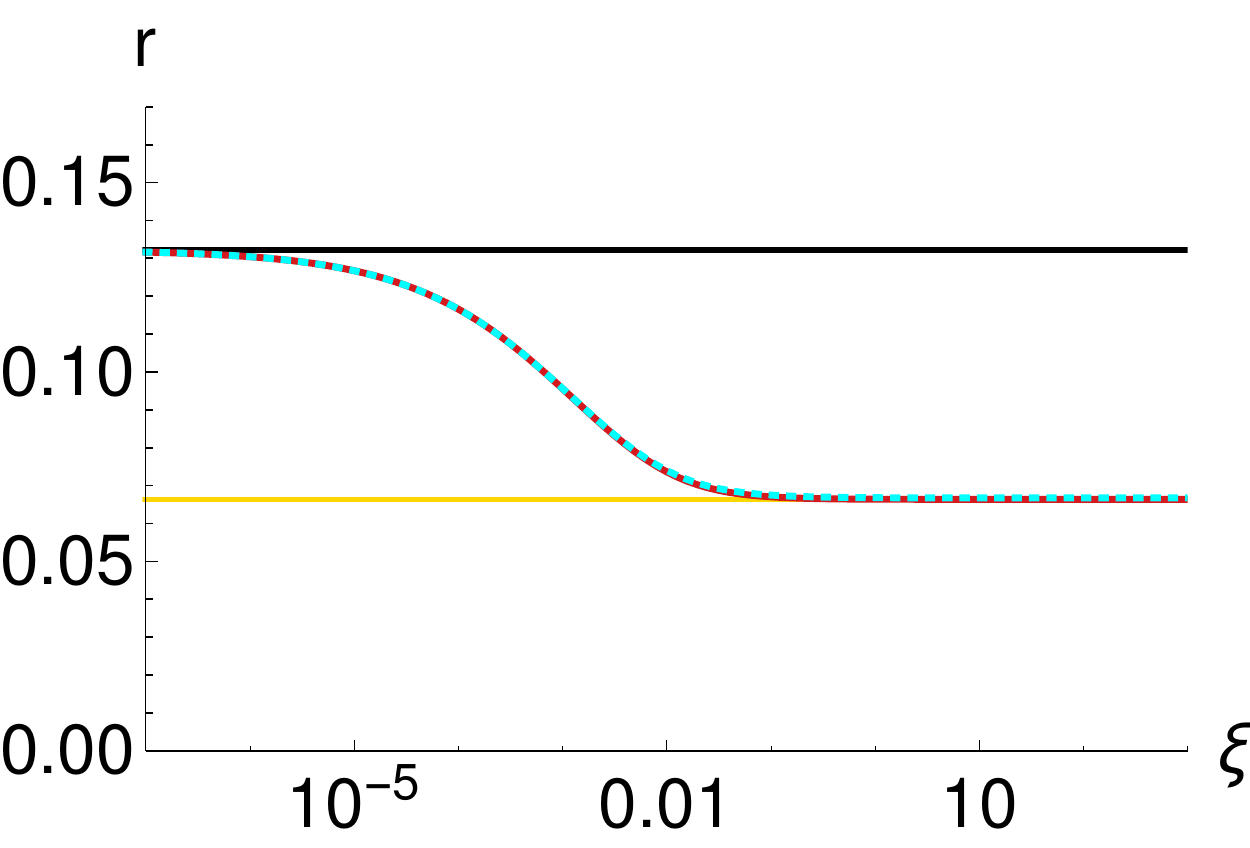}\\
\quad\\
\quad\\
\hspace{-0.3cm}\includegraphics[width=0.45 \textwidth]{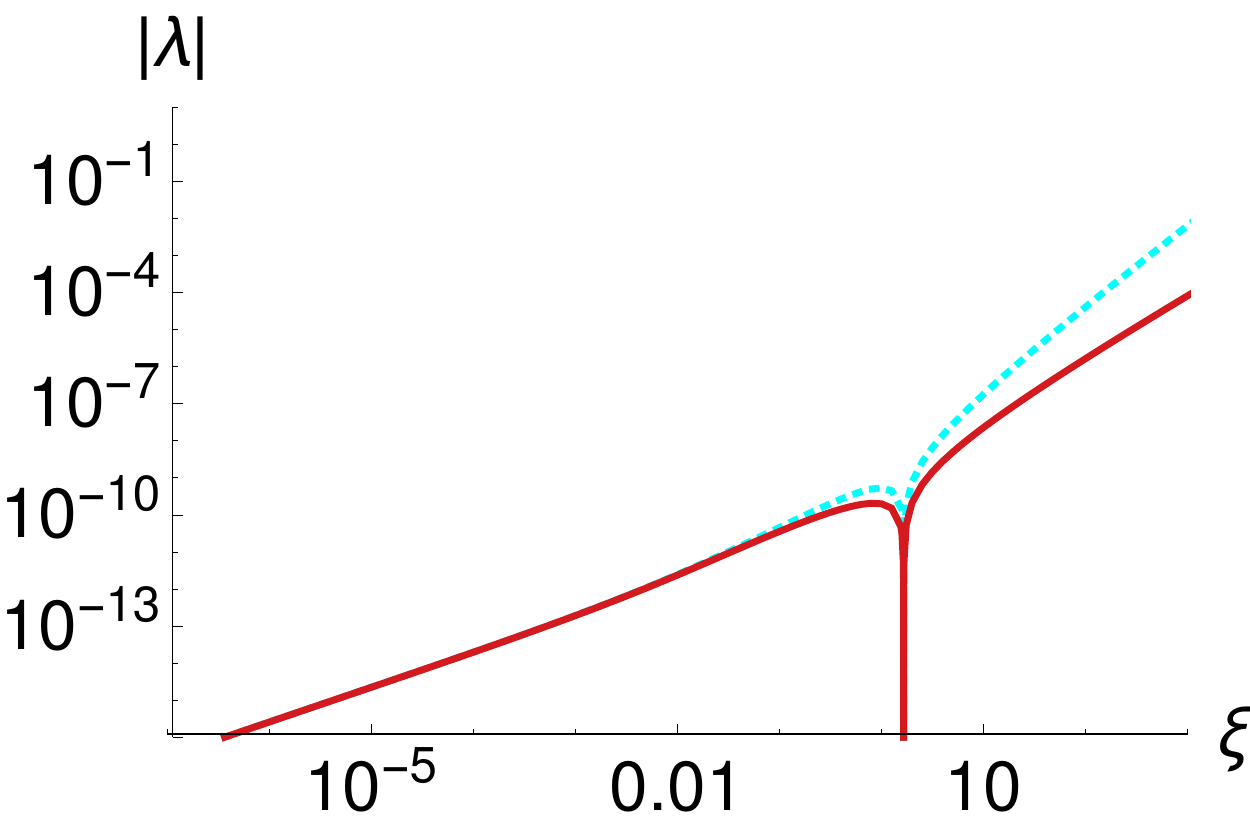}\\
\quad\\
\quad\\
\caption{$r$ vs. $n_s$ (upper panel), $r$ vs. $\xi$ (central panel) and $|\lambda|$ vs. $\xi$ (lower panel) for $N = 60$ $e$-folds in the metric (cyan, dashed) and Palatini formulation (red) in the induced gravity scenario. For reference, we also plot the predictions of quadratic (black), linear (yellow) and Starobinsky (orange) inflation in metric gravity.  The light blue areas present the 1 and 2$\sigma$ constraints from the BICEP2/Keck data \cite{Array:2015xqh}. 
    }
   \label{Fig:r:vs:ns:ind}
  \end{center}
\end{figure}
The corresponding results are given in Fig.  \ref{Fig:r:vs:ns:ind} where we depict $r$ vs. $n_s$ (upper panel), $r$ vs. $\xi$ (central panel) and the absolute value of $\lambda$ ($|\lambda|$) vs. $\xi$ (lower panel) for $N = 60$ $e$-folds in the metric (cyan, dashed) and Palatini formulation (red) in the induced gravity scenario. For reference, we also plot the predictions of quadratic (black), linear (yellow) and Starobinsky (orange) inflation in metric gravity.  The light blue areas present the 1 and 2$\sigma$ constraints from the BICEP2/Keck data \cite{Array:2015xqh}.
As already shown in \cite{Racioppi:2017spw}, the predictions are confined in between those of quadratic and linear inflation and it is essentially impossible to distinguish between the two formulations without taking into account reheating. For more details, we refer  the reader to \cite{Racioppi:2017spw}.
In addition to the study of \cite{Racioppi:2017spw}, we show here explicitly that the model remains perturbative  until the linear limit in both the formulations. The cusp in the $|\lambda|$ vs. $\xi$ plot at $\xi=\sqrt e$ is an effect of the logarithmic scale. Such a value corresponds to the case $\lambda =0$ (and therefore $\delta = \infty$) and it means that our parametrization is not a good one for such a value of $\xi$. However, it can be shown that the model is perfectly consistent, since the product $\lambda\delta = \beta_\lambda$ remains finite.


\section{Conclusions}
\label{conclusions}

We performed an analysis of models of quartic inflation where the inflaton field $\phi$ is subject to relevant loop corrections and it is coupled non-minimally to gravity. We considered two types of quadratic non-minimal couplings: Higgs-inflation-like and induced gravity one. For both of these, we studied the predictions of two different formulations of gravity: {\it metric} and {\it Palatini}. We showed that in all the cases studied the famous Starobinsky attractor is lost, having been replaced by linear inflation. We stress that the existence of such property is {\it universal} i.e. independent of the underlying theory of gravity.

\section*{Acknowledgements}
The author thanks Martti Raidal, Luca Marzola and Micha{\l}~Artymowski for useful discussions.
This work was supported by the Estonian Research Council grants IUT23-6, PUT1026, and by the ERDF Centre of Excellence project TK133.

\bibliography{references}

\begin{thebibliography}{86}%
\makeatletter
\providecommand \@ifxundefined [1]{%
 \@ifx{#1\undefined}
}%
\providecommand \@ifnum [1]{%
 \ifnum #1\expandafter \@firstoftwo
 \else \expandafter \@secondoftwo
 \fi
}%
\providecommand \@ifx [1]{%
 \ifx #1\expandafter \@firstoftwo
 \else \expandafter \@secondoftwo
 \fi
}%
\providecommand \natexlab [1]{#1}%
\providecommand \enquote  [1]{``#1''}%
\providecommand \bibnamefont  [1]{#1}%
\providecommand \bibfnamefont [1]{#1}%
\providecommand \citenamefont [1]{#1}%
\providecommand \href@noop [0]{\@secondoftwo}%
\providecommand \href [0]{\begingroup \@sanitize@url \@href}%
\providecommand \@href[1]{\@@startlink{#1}\@@href}%
\providecommand \@@href[1]{\endgroup#1\@@endlink}%
\providecommand \@sanitize@url [0]{\catcode `\\12\catcode `\$12\catcode
  `\&12\catcode `\#12\catcode `\^12\catcode `\_12\catcode `\%12\relax}%
\providecommand \@@startlink[1]{}%
\providecommand \@@endlink[0]{}%
\providecommand \url  [0]{\begingroup\@sanitize@url \@url }%
\providecommand \@url [1]{\endgroup\@href {#1}{\urlprefix }}%
\providecommand \urlprefix  [0]{URL }%
\providecommand \Eprint [0]{\href }%
\providecommand \doibase [0]{http://dx.doi.org/}%
\providecommand \selectlanguage [0]{\@gobble}%
\providecommand \bibinfo  [0]{\@secondoftwo}%
\providecommand \bibfield  [0]{\@secondoftwo}%
\providecommand \translation [1]{[#1]}%
\providecommand \BibitemOpen [0]{}%
\providecommand \bibitemStop [0]{}%
\providecommand \bibitemNoStop [0]{.\EOS\space}%
\providecommand \EOS [0]{\spacefactor3000\relax}%
\providecommand \BibitemShut  [1]{\csname bibitem#1\endcsname}%
\let\auto@bib@innerbib\@empty
\bibitem [{\citenamefont {Starobinsky}(1980)}]{Starobinsky:1980te}%
  \BibitemOpen
  \bibfield  {author} {\bibinfo {author} {\bibfnamefont {A.~A.}\ \bibnamefont
  {Starobinsky}},\ }\href {\doibase 10.1016/0370-2693(80)90670-X} {\bibfield
  {journal} {\bibinfo  {journal} {Phys. Lett.}\ }\textbf {\bibinfo {volume}
  {B91}},\ \bibinfo {pages} {99} (\bibinfo {year} {1980})}\BibitemShut
  {NoStop}%
\bibitem [{\citenamefont {Guth}(1981)}]{Guth:1980zm}%
  \BibitemOpen
  \bibfield  {author} {\bibinfo {author} {\bibfnamefont {A.~H.}\ \bibnamefont
  {Guth}},\ }\href {\doibase 10.1103/PhysRevD.23.347} {\bibfield  {journal}
  {\bibinfo  {journal} {Phys.Rev.}\ }\textbf {\bibinfo {volume} {D23}},\
  \bibinfo {pages} {347} (\bibinfo {year} {1981})}\BibitemShut {NoStop}%
\bibitem [{\citenamefont {Linde}(1982)}]{Linde:1981mu}%
  \BibitemOpen
  \bibfield  {author} {\bibinfo {author} {\bibfnamefont {A.~D.}\ \bibnamefont
  {Linde}},\ }\href {\doibase 10.1016/0370-2693(82)91219-9} {\bibfield
  {journal} {\bibinfo  {journal} {Phys.Lett.}\ }\textbf {\bibinfo {volume}
  {B108}},\ \bibinfo {pages} {389} (\bibinfo {year} {1982})}\BibitemShut
  {NoStop}%
\bibitem [{\citenamefont {Albrecht}\ and\ \citenamefont
  {Steinhardt}(1982)}]{Albrecht:1982wi}%
  \BibitemOpen
  \bibfield  {author} {\bibinfo {author} {\bibfnamefont {A.}~\bibnamefont
  {Albrecht}}\ and\ \bibinfo {author} {\bibfnamefont {P.~J.}\ \bibnamefont
  {Steinhardt}},\ }\href {\doibase 10.1103/PhysRevLett.48.1220} {\bibfield
  {journal} {\bibinfo  {journal} {Phys.Rev.Lett.}\ }\textbf {\bibinfo {volume}
  {48}},\ \bibinfo {pages} {1220} (\bibinfo {year} {1982})}\BibitemShut
  {NoStop}%
\bibitem [{\citenamefont {Ade}\ \emph {et~al.}(2015)\citenamefont {Ade} \emph
  {et~al.}}]{Ade:2015tva}%
  \BibitemOpen
  \bibfield  {author} {\bibinfo {author} {\bibfnamefont {P.~A.~R.}\
  \bibnamefont {Ade}} \emph {et~al.} (\bibinfo {collaboration} {BICEP2,
  Planck}),\ }\href {\doibase 10.1103/PhysRevLett.114.101301} {\bibfield
  {journal} {\bibinfo  {journal} {Phys. Rev. Lett.}\ }\textbf {\bibinfo
  {volume} {114}},\ \bibinfo {pages} {101301} (\bibinfo {year} {2015})},\
  \Eprint {http://arxiv.org/abs/1502.00612} {arXiv:1502.00612 [astro-ph.CO]}
  \BibitemShut {NoStop}%
\bibitem [{\citenamefont {Ade}\ \emph {et~al.}(2016{\natexlab{a}})\citenamefont
  {Ade} \emph {et~al.}}]{Ade:2015xua}%
  \BibitemOpen
  \bibfield  {author} {\bibinfo {author} {\bibfnamefont {P.~A.~R.}\
  \bibnamefont {Ade}} \emph {et~al.} (\bibinfo {collaboration} {Planck}),\
  }\href {\doibase 10.1051/0004-6361/201525830} {\bibfield  {journal} {\bibinfo
   {journal} {Astron. Astrophys.}\ }\textbf {\bibinfo {volume} {594}},\
  \bibinfo {pages} {A13} (\bibinfo {year} {2016}{\natexlab{a}})},\ \Eprint
  {http://arxiv.org/abs/1502.01589} {arXiv:1502.01589 [astro-ph.CO]}
  \BibitemShut {NoStop}%
\bibitem [{\citenamefont {Ade}\ \emph {et~al.}(2016{\natexlab{b}})\citenamefont
  {Ade} \emph {et~al.}}]{Ade:2015lrj}%
  \BibitemOpen
  \bibfield  {author} {\bibinfo {author} {\bibfnamefont {P.~A.~R.}\
  \bibnamefont {Ade}} \emph {et~al.} (\bibinfo {collaboration} {Planck}),\
  }\href {\doibase 10.1051/0004-6361/201525898} {\bibfield  {journal} {\bibinfo
   {journal} {Astron. Astrophys.}\ }\textbf {\bibinfo {volume} {594}},\
  \bibinfo {pages} {A20} (\bibinfo {year} {2016}{\natexlab{b}})},\ \Eprint
  {http://arxiv.org/abs/1502.02114} {arXiv:1502.02114 [astro-ph.CO]}
  \BibitemShut {NoStop}%
\bibitem [{\citenamefont {Ade}\ \emph {et~al.}(2016{\natexlab{c}})\citenamefont
  {Ade} \emph {et~al.}}]{Array:2015xqh}%
  \BibitemOpen
  \bibfield  {author} {\bibinfo {author} {\bibfnamefont {P.~A.~R.}\
  \bibnamefont {Ade}} \emph {et~al.} (\bibinfo {collaboration} {BICEP2, Keck
  Array}),\ }\href {\doibase 10.1103/PhysRevLett.116.031302} {\bibfield
  {journal} {\bibinfo  {journal} {Phys. Rev. Lett.}\ }\textbf {\bibinfo
  {volume} {116}},\ \bibinfo {pages} {031302} (\bibinfo {year}
  {2016}{\natexlab{c}})},\ \Eprint {http://arxiv.org/abs/1510.09217}
  {arXiv:1510.09217 [astro-ph.CO]} \BibitemShut {NoStop}%
\bibitem [{\citenamefont {Boubekeur}\ and\ \citenamefont
  {Lyth}(2005)}]{Boubekeur:2005zm}%
  \BibitemOpen
  \bibfield  {author} {\bibinfo {author} {\bibfnamefont {L.}~\bibnamefont
  {Boubekeur}}\ and\ \bibinfo {author} {\bibfnamefont {D.~H.}\ \bibnamefont
  {Lyth}},\ }\href {\doibase 10.1088/1475-7516/2005/07/010} {\bibfield
  {journal} {\bibinfo  {journal} {JCAP}\ }\textbf {\bibinfo {volume} {0507}},\
  \bibinfo {pages} {010} (\bibinfo {year} {2005})},\ \Eprint
  {http://arxiv.org/abs/hep-ph/0502047} {arXiv:hep-ph/0502047 [hep-ph]}
  \BibitemShut {NoStop}%
\bibitem [{\citenamefont {McAllister}\ \emph {et~al.}(2010)\citenamefont
  {McAllister}, \citenamefont {Silverstein},\ and\ \citenamefont
  {Westphal}}]{McAllister:2008hb}%
  \BibitemOpen
  \bibfield  {author} {\bibinfo {author} {\bibfnamefont {L.}~\bibnamefont
  {McAllister}}, \bibinfo {author} {\bibfnamefont {E.}~\bibnamefont
  {Silverstein}}, \ and\ \bibinfo {author} {\bibfnamefont {A.}~\bibnamefont
  {Westphal}},\ }\href {\doibase 10.1103/PhysRevD.82.046003} {\bibfield
  {journal} {\bibinfo  {journal} {Phys. Rev.}\ }\textbf {\bibinfo {volume}
  {D82}},\ \bibinfo {pages} {046003} (\bibinfo {year} {2010})},\ \Eprint
  {http://arxiv.org/abs/0808.0706} {arXiv:0808.0706 [hep-th]} \BibitemShut
  {NoStop}%
\bibitem [{\citenamefont {Iso}\ \emph {et~al.}(2015)\citenamefont {Iso},
  \citenamefont {Kohri},\ and\ \citenamefont {Shimada}}]{Iso:2014gka}%
  \BibitemOpen
  \bibfield  {author} {\bibinfo {author} {\bibfnamefont {S.}~\bibnamefont
  {Iso}}, \bibinfo {author} {\bibfnamefont {K.}~\bibnamefont {Kohri}}, \ and\
  \bibinfo {author} {\bibfnamefont {K.}~\bibnamefont {Shimada}},\ }\href
  {\doibase 10.1103/PhysRevD.91.044006} {\bibfield  {journal} {\bibinfo
  {journal} {Phys. Rev.}\ }\textbf {\bibinfo {volume} {D91}},\ \bibinfo {pages}
  {044006} (\bibinfo {year} {2015})},\ \Eprint {http://arxiv.org/abs/1408.2339}
  {arXiv:1408.2339 [hep-ph]} \BibitemShut {NoStop}%
\bibitem [{\citenamefont {Kannike}\ \emph {et~al.}(2016)\citenamefont
  {Kannike}, \citenamefont {Racioppi},\ and\ \citenamefont
  {Raidal}}]{Kannike:2015kda}%
  \BibitemOpen
  \bibfield  {author} {\bibinfo {author} {\bibfnamefont {K.}~\bibnamefont
  {Kannike}}, \bibinfo {author} {\bibfnamefont {A.}~\bibnamefont {Racioppi}}, \
  and\ \bibinfo {author} {\bibfnamefont {M.}~\bibnamefont {Raidal}},\ }\href
  {\doibase 10.1007/JHEP01(2016)035} {\bibfield  {journal} {\bibinfo  {journal}
  {JHEP}\ }\textbf {\bibinfo {volume} {01}},\ \bibinfo {pages} {035} (\bibinfo
  {year} {2016})},\ \Eprint {http://arxiv.org/abs/1509.05423} {arXiv:1509.05423
  [hep-ph]} \BibitemShut {NoStop}%
\bibitem [{\citenamefont {Rinaldi}\ \emph {et~al.}(2016)\citenamefont
  {Rinaldi}, \citenamefont {Vanzo}, \citenamefont {Zerbini},\ and\
  \citenamefont {Venturi}}]{Rinaldi:2015yoa}%
  \BibitemOpen
  \bibfield  {author} {\bibinfo {author} {\bibfnamefont {M.}~\bibnamefont
  {Rinaldi}}, \bibinfo {author} {\bibfnamefont {L.}~\bibnamefont {Vanzo}},
  \bibinfo {author} {\bibfnamefont {S.}~\bibnamefont {Zerbini}}, \ and\
  \bibinfo {author} {\bibfnamefont {G.}~\bibnamefont {Venturi}},\ }\href
  {\doibase 10.1103/PhysRevD.93.024040} {\bibfield  {journal} {\bibinfo
  {journal} {Phys. Rev.}\ }\textbf {\bibinfo {volume} {D93}},\ \bibinfo {pages}
  {024040} (\bibinfo {year} {2016})},\ \Eprint
  {http://arxiv.org/abs/1505.03386} {arXiv:1505.03386 [hep-th]} \BibitemShut
  {NoStop}%
\bibitem [{\citenamefont {Barrie}\ \emph {et~al.}(2016)\citenamefont {Barrie},
  \citenamefont {Kobakhidze},\ and\ \citenamefont {Liang}}]{Barrie:2016rnv}%
  \BibitemOpen
  \bibfield  {author} {\bibinfo {author} {\bibfnamefont {N.~D.}\ \bibnamefont
  {Barrie}}, \bibinfo {author} {\bibfnamefont {A.}~\bibnamefont {Kobakhidze}},
  \ and\ \bibinfo {author} {\bibfnamefont {S.}~\bibnamefont {Liang}},\ }\href
  {\doibase 10.1016/j.physletb.2016.03.056} {\bibfield  {journal} {\bibinfo
  {journal} {Phys. Lett.}\ }\textbf {\bibinfo {volume} {B756}},\ \bibinfo
  {pages} {390} (\bibinfo {year} {2016})},\ \Eprint
  {http://arxiv.org/abs/1602.04901} {arXiv:1602.04901 [gr-qc]} \BibitemShut
  {NoStop}%
\bibitem [{\citenamefont {Artymowski}\ and\ \citenamefont
  {Racioppi}(2017)}]{Artymowski:2016dlz}%
  \BibitemOpen
  \bibfield  {author} {\bibinfo {author} {\bibfnamefont {M.}~\bibnamefont
  {Artymowski}}\ and\ \bibinfo {author} {\bibfnamefont {A.}~\bibnamefont
  {Racioppi}},\ }\href {\doibase 10.1088/1475-7516/2017/04/007} {\bibfield
  {journal} {\bibinfo  {journal} {JCAP}\ }\textbf {\bibinfo {volume} {1704}},\
  \bibinfo {pages} {007} (\bibinfo {year} {2017})},\ \Eprint
  {http://arxiv.org/abs/1610.09120} {arXiv:1610.09120 [astro-ph.CO]}
  \BibitemShut {NoStop}%
\bibitem [{\citenamefont {Racioppi}(2017)}]{Racioppi:2017spw}%
  \BibitemOpen
  \bibfield  {author} {\bibinfo {author} {\bibfnamefont {A.}~\bibnamefont
  {Racioppi}},\ }\href {\doibase 10.1088/1475-7516/2017/12/041} {\bibfield
  {journal} {\bibinfo  {journal} {JCAP}\ }\textbf {\bibinfo {volume} {1712}},\
  \bibinfo {pages} {041} (\bibinfo {year} {2017})},\ \Eprint
  {http://arxiv.org/abs/1710.04853} {arXiv:1710.04853 [astro-ph.CO]}
  \BibitemShut {NoStop}%
\bibitem [{\citenamefont {Karam}\ \emph
  {et~al.}(2017{\natexlab{a}})\citenamefont {Karam}, \citenamefont {Marzola},
  \citenamefont {Pappas}, \citenamefont {Racioppi},\ and\ \citenamefont
  {Tamvakis}}]{Karam:2017rpw}%
  \BibitemOpen
  \bibfield  {author} {\bibinfo {author} {\bibfnamefont {A.}~\bibnamefont
  {Karam}}, \bibinfo {author} {\bibfnamefont {L.}~\bibnamefont {Marzola}},
  \bibinfo {author} {\bibfnamefont {T.}~\bibnamefont {Pappas}}, \bibinfo
  {author} {\bibfnamefont {A.}~\bibnamefont {Racioppi}}, \ and\ \bibinfo
  {author} {\bibfnamefont {K.}~\bibnamefont {Tamvakis}},\ }\href@noop {} {\
  (\bibinfo {year} {2017}{\natexlab{a}})},\ \Eprint
  {http://arxiv.org/abs/1711.09861} {arXiv:1711.09861 [astro-ph.CO]}
  \BibitemShut {NoStop}%
\bibitem [{\citenamefont {Futamase}\ and\ \citenamefont
  {Maeda}(1989)}]{Futamase:1987ua}%
  \BibitemOpen
  \bibfield  {author} {\bibinfo {author} {\bibfnamefont {T.}~\bibnamefont
  {Futamase}}\ and\ \bibinfo {author} {\bibfnamefont {K.-i.}\ \bibnamefont
  {Maeda}},\ }\href {\doibase 10.1103/PhysRevD.39.399} {\bibfield  {journal}
  {\bibinfo  {journal} {Phys. Rev.}\ }\textbf {\bibinfo {volume} {D39}},\
  \bibinfo {pages} {399} (\bibinfo {year} {1989})}\BibitemShut {NoStop}%
\bibitem [{\citenamefont {Salopek}\ \emph {et~al.}(1989)\citenamefont
  {Salopek}, \citenamefont {Bond},\ and\ \citenamefont
  {Bardeen}}]{Salopek:1988qh}%
  \BibitemOpen
  \bibfield  {author} {\bibinfo {author} {\bibfnamefont {D.~S.}\ \bibnamefont
  {Salopek}}, \bibinfo {author} {\bibfnamefont {J.~R.}\ \bibnamefont {Bond}}, \
  and\ \bibinfo {author} {\bibfnamefont {J.~M.}\ \bibnamefont {Bardeen}},\
  }\href {\doibase 10.1103/PhysRevD.40.1753} {\bibfield  {journal} {\bibinfo
  {journal} {Phys. Rev.}\ }\textbf {\bibinfo {volume} {D40}},\ \bibinfo {pages}
  {1753} (\bibinfo {year} {1989})}\BibitemShut {NoStop}%
\bibitem [{\citenamefont {Fakir}\ and\ \citenamefont
  {Unruh}(1990)}]{Fakir:1990eg}%
  \BibitemOpen
  \bibfield  {author} {\bibinfo {author} {\bibfnamefont {R.}~\bibnamefont
  {Fakir}}\ and\ \bibinfo {author} {\bibfnamefont {W.~G.}\ \bibnamefont
  {Unruh}},\ }\href {\doibase 10.1103/PhysRevD.41.1783} {\bibfield  {journal}
  {\bibinfo  {journal} {Phys. Rev.}\ }\textbf {\bibinfo {volume} {D41}},\
  \bibinfo {pages} {1783} (\bibinfo {year} {1990})}\BibitemShut {NoStop}%
\bibitem [{\citenamefont {Amendola}\ \emph {et~al.}(1990)\citenamefont
  {Amendola}, \citenamefont {Litterio},\ and\ \citenamefont
  {Occhionero}}]{Amendola:1990nn}%
  \BibitemOpen
  \bibfield  {author} {\bibinfo {author} {\bibfnamefont {L.}~\bibnamefont
  {Amendola}}, \bibinfo {author} {\bibfnamefont {M.}~\bibnamefont {Litterio}},
  \ and\ \bibinfo {author} {\bibfnamefont {F.}~\bibnamefont {Occhionero}},\
  }\href {\doibase 10.1142/S0217751X90001653} {\bibfield  {journal} {\bibinfo
  {journal} {Int. J. Mod. Phys.}\ }\textbf {\bibinfo {volume} {A5}},\ \bibinfo
  {pages} {3861} (\bibinfo {year} {1990})}\BibitemShut {NoStop}%
\bibitem [{\citenamefont {Kaiser}(1995)}]{Kaiser:1994vs}%
  \BibitemOpen
  \bibfield  {author} {\bibinfo {author} {\bibfnamefont {D.~I.}\ \bibnamefont
  {Kaiser}},\ }\href {\doibase 10.1103/PhysRevD.52.4295} {\bibfield  {journal}
  {\bibinfo  {journal} {Phys. Rev.}\ }\textbf {\bibinfo {volume} {D52}},\
  \bibinfo {pages} {4295} (\bibinfo {year} {1995})},\ \Eprint
  {http://arxiv.org/abs/astro-ph/9408044} {arXiv:astro-ph/9408044 [astro-ph]}
  \BibitemShut {NoStop}%
\bibitem [{\citenamefont {Bezrukov}\ and\ \citenamefont
  {Shaposhnikov}(2008)}]{Bezrukov:2007ep}%
  \BibitemOpen
  \bibfield  {author} {\bibinfo {author} {\bibfnamefont {F.~L.}\ \bibnamefont
  {Bezrukov}}\ and\ \bibinfo {author} {\bibfnamefont {M.}~\bibnamefont
  {Shaposhnikov}},\ }\href {\doibase 10.1016/j.physletb.2007.11.072} {\bibfield
   {journal} {\bibinfo  {journal} {Phys. Lett.}\ }\textbf {\bibinfo {volume}
  {B659}},\ \bibinfo {pages} {703} (\bibinfo {year} {2008})},\ \Eprint
  {http://arxiv.org/abs/0710.3755} {arXiv:0710.3755 [hep-th]} \BibitemShut
  {NoStop}%
\bibitem [{\citenamefont {Bauer}\ and\ \citenamefont
  {Demir}(2008)}]{Bauer:2008zj}%
  \BibitemOpen
  \bibfield  {author} {\bibinfo {author} {\bibfnamefont {F.}~\bibnamefont
  {Bauer}}\ and\ \bibinfo {author} {\bibfnamefont {D.~A.}\ \bibnamefont
  {Demir}},\ }\href {\doibase 10.1016/j.physletb.2008.06.014} {\bibfield
  {journal} {\bibinfo  {journal} {Phys. Lett.}\ }\textbf {\bibinfo {volume}
  {B665}},\ \bibinfo {pages} {222} (\bibinfo {year} {2008})},\ \Eprint
  {http://arxiv.org/abs/0803.2664} {arXiv:0803.2664 [hep-ph]} \BibitemShut
  {NoStop}%
\bibitem [{\citenamefont {Park}\ and\ \citenamefont
  {Yamaguchi}(2008)}]{Park:2008hz}%
  \BibitemOpen
  \bibfield  {author} {\bibinfo {author} {\bibfnamefont {S.~C.}\ \bibnamefont
  {Park}}\ and\ \bibinfo {author} {\bibfnamefont {S.}~\bibnamefont
  {Yamaguchi}},\ }\href {\doibase 10.1088/1475-7516/2008/08/009} {\bibfield
  {journal} {\bibinfo  {journal} {JCAP}\ }\textbf {\bibinfo {volume} {0808}},\
  \bibinfo {pages} {009} (\bibinfo {year} {2008})},\ \Eprint
  {http://arxiv.org/abs/0801.1722} {arXiv:0801.1722 [hep-ph]} \BibitemShut
  {NoStop}%
\bibitem [{\citenamefont {Linde}\ \emph {et~al.}(2011)\citenamefont {Linde},
  \citenamefont {Noorbala},\ and\ \citenamefont {Westphal}}]{Linde:2011nh}%
  \BibitemOpen
  \bibfield  {author} {\bibinfo {author} {\bibfnamefont {A.}~\bibnamefont
  {Linde}}, \bibinfo {author} {\bibfnamefont {M.}~\bibnamefont {Noorbala}}, \
  and\ \bibinfo {author} {\bibfnamefont {A.}~\bibnamefont {Westphal}},\ }\href
  {\doibase 10.1088/1475-7516/2011/03/013} {\bibfield  {journal} {\bibinfo
  {journal} {JCAP}\ }\textbf {\bibinfo {volume} {1103}},\ \bibinfo {pages}
  {013} (\bibinfo {year} {2011})},\ \Eprint {http://arxiv.org/abs/1101.2652}
  {arXiv:1101.2652 [hep-th]} \BibitemShut {NoStop}%
\bibitem [{\citenamefont {Kaiser}\ and\ \citenamefont
  {Sfakianakis}(2014)}]{Kaiser:2013sna}%
  \BibitemOpen
  \bibfield  {author} {\bibinfo {author} {\bibfnamefont {D.~I.}\ \bibnamefont
  {Kaiser}}\ and\ \bibinfo {author} {\bibfnamefont {E.~I.}\ \bibnamefont
  {Sfakianakis}},\ }\href {\doibase 10.1103/PhysRevLett.112.011302} {\bibfield
  {journal} {\bibinfo  {journal} {Phys. Rev. Lett.}\ }\textbf {\bibinfo
  {volume} {112}},\ \bibinfo {pages} {011302} (\bibinfo {year} {2014})},\
  \Eprint {http://arxiv.org/abs/1304.0363} {arXiv:1304.0363 [astro-ph.CO]}
  \BibitemShut {NoStop}%
\bibitem [{\citenamefont {Kallosh}\ and\ \citenamefont
  {Linde}(2013{\natexlab{a}})}]{Kallosh:2013maa}%
  \BibitemOpen
  \bibfield  {author} {\bibinfo {author} {\bibfnamefont {R.}~\bibnamefont
  {Kallosh}}\ and\ \bibinfo {author} {\bibfnamefont {A.}~\bibnamefont
  {Linde}},\ }\href {\doibase 10.1088/1475-7516/2013/10/033} {\bibfield
  {journal} {\bibinfo  {journal} {JCAP}\ }\textbf {\bibinfo {volume} {1310}},\
  \bibinfo {pages} {033} (\bibinfo {year} {2013}{\natexlab{a}})},\ \Eprint
  {http://arxiv.org/abs/1307.7938} {arXiv:1307.7938 [hep-th]} \BibitemShut
  {NoStop}%
\bibitem [{\citenamefont {Kallosh}\ and\ \citenamefont
  {Linde}(2013{\natexlab{b}})}]{Kallosh:2013daa}%
  \BibitemOpen
  \bibfield  {author} {\bibinfo {author} {\bibfnamefont {R.}~\bibnamefont
  {Kallosh}}\ and\ \bibinfo {author} {\bibfnamefont {A.}~\bibnamefont
  {Linde}},\ }\href {\doibase 10.1088/1475-7516/2013/12/006} {\bibfield
  {journal} {\bibinfo  {journal} {JCAP}\ }\textbf {\bibinfo {volume} {1312}},\
  \bibinfo {pages} {006} (\bibinfo {year} {2013}{\natexlab{b}})},\ \Eprint
  {http://arxiv.org/abs/1309.2015} {arXiv:1309.2015 [hep-th]} \BibitemShut
  {NoStop}%
\bibitem [{\citenamefont {Kallosh}\ \emph {et~al.}(2014)\citenamefont
  {Kallosh}, \citenamefont {Linde},\ and\ \citenamefont
  {Roest}}]{Kallosh:2013tua}%
  \BibitemOpen
  \bibfield  {author} {\bibinfo {author} {\bibfnamefont {R.}~\bibnamefont
  {Kallosh}}, \bibinfo {author} {\bibfnamefont {A.}~\bibnamefont {Linde}}, \
  and\ \bibinfo {author} {\bibfnamefont {D.}~\bibnamefont {Roest}},\ }\href
  {\doibase 10.1103/PhysRevLett.112.011303} {\bibfield  {journal} {\bibinfo
  {journal} {Phys. Rev. Lett.}\ }\textbf {\bibinfo {volume} {112}},\ \bibinfo
  {pages} {011303} (\bibinfo {year} {2014})},\ \Eprint
  {http://arxiv.org/abs/1310.3950} {arXiv:1310.3950 [hep-th]} \BibitemShut
  {NoStop}%
\bibitem [{\citenamefont {Rubio}\ and\ \citenamefont
  {Shaposhnikov}(2014)}]{Rubio:2014wta}%
  \BibitemOpen
  \bibfield  {author} {\bibinfo {author} {\bibfnamefont {J.}~\bibnamefont
  {Rubio}}\ and\ \bibinfo {author} {\bibfnamefont {M.}~\bibnamefont
  {Shaposhnikov}},\ }\href {\doibase 10.1103/PhysRevD.90.027307} {\bibfield
  {journal} {\bibinfo  {journal} {Phys. Rev.}\ }\textbf {\bibinfo {volume}
  {D90}},\ \bibinfo {pages} {027307} (\bibinfo {year} {2014})},\ \Eprint
  {http://arxiv.org/abs/1406.5182} {arXiv:1406.5182 [hep-ph]} \BibitemShut
  {NoStop}%
\bibitem [{\citenamefont {Csaki}\ \emph {et~al.}(2014)\citenamefont {Csaki},
  \citenamefont {Kaloper}, \citenamefont {Serra},\ and\ \citenamefont
  {Terning}}]{Csaki:2014bua}%
  \BibitemOpen
  \bibfield  {author} {\bibinfo {author} {\bibfnamefont {C.}~\bibnamefont
  {Csaki}}, \bibinfo {author} {\bibfnamefont {N.}~\bibnamefont {Kaloper}},
  \bibinfo {author} {\bibfnamefont {J.}~\bibnamefont {Serra}}, \ and\ \bibinfo
  {author} {\bibfnamefont {J.}~\bibnamefont {Terning}},\ }\href {\doibase
  10.1103/PhysRevLett.113.161302} {\bibfield  {journal} {\bibinfo  {journal}
  {Phys. Rev. Lett.}\ }\textbf {\bibinfo {volume} {113}},\ \bibinfo {pages}
  {161302} (\bibinfo {year} {2014})},\ \Eprint {http://arxiv.org/abs/1406.5192}
  {arXiv:1406.5192 [hep-th]} \BibitemShut {NoStop}%
\bibitem [{\citenamefont {Galante}\ \emph {et~al.}(2015)\citenamefont
  {Galante}, \citenamefont {Kallosh}, \citenamefont {Linde},\ and\
  \citenamefont {Roest}}]{Galante:2014ifa}%
  \BibitemOpen
  \bibfield  {author} {\bibinfo {author} {\bibfnamefont {M.}~\bibnamefont
  {Galante}}, \bibinfo {author} {\bibfnamefont {R.}~\bibnamefont {Kallosh}},
  \bibinfo {author} {\bibfnamefont {A.}~\bibnamefont {Linde}}, \ and\ \bibinfo
  {author} {\bibfnamefont {D.}~\bibnamefont {Roest}},\ }\href {\doibase
  10.1103/PhysRevLett.114.141302} {\bibfield  {journal} {\bibinfo  {journal}
  {Phys. Rev. Lett.}\ }\textbf {\bibinfo {volume} {114}},\ \bibinfo {pages}
  {141302} (\bibinfo {year} {2015})},\ \Eprint {http://arxiv.org/abs/1412.3797}
  {arXiv:1412.3797 [hep-th]} \BibitemShut {NoStop}%
\bibitem [{\citenamefont {Chiba}\ and\ \citenamefont
  {Kohri}(2015)}]{Chiba:2014sva}%
  \BibitemOpen
  \bibfield  {author} {\bibinfo {author} {\bibfnamefont {T.}~\bibnamefont
  {Chiba}}\ and\ \bibinfo {author} {\bibfnamefont {K.}~\bibnamefont {Kohri}},\
  }\href {\doibase 10.1093/ptep/ptv007} {\bibfield  {journal} {\bibinfo
  {journal} {PTEP}\ }\textbf {\bibinfo {volume} {2015}},\ \bibinfo {pages}
  {023E01} (\bibinfo {year} {2015})},\ \Eprint {http://arxiv.org/abs/1411.7104}
  {arXiv:1411.7104 [astro-ph.CO]} \BibitemShut {NoStop}%
\bibitem [{\citenamefont {Boubekeur}\ \emph {et~al.}(2015)\citenamefont
  {Boubekeur}, \citenamefont {Giusarma}, \citenamefont {Mena},\ and\
  \citenamefont {Ramírez}}]{Boubekeur:2015xza}%
  \BibitemOpen
  \bibfield  {author} {\bibinfo {author} {\bibfnamefont {L.}~\bibnamefont
  {Boubekeur}}, \bibinfo {author} {\bibfnamefont {E.}~\bibnamefont {Giusarma}},
  \bibinfo {author} {\bibfnamefont {O.}~\bibnamefont {Mena}}, \ and\ \bibinfo
  {author} {\bibfnamefont {H.}~\bibnamefont {Ramírez}},\ }\href {\doibase
  10.1103/PhysRevD.91.103004} {\bibfield  {journal} {\bibinfo  {journal} {Phys.
  Rev.}\ }\textbf {\bibinfo {volume} {D91}},\ \bibinfo {pages} {103004}
  (\bibinfo {year} {2015})},\ \Eprint {http://arxiv.org/abs/1502.05193}
  {arXiv:1502.05193 [astro-ph.CO]} \BibitemShut {NoStop}%
\bibitem [{\citenamefont {Pieroni}(2016)}]{Pieroni:2015cma}%
  \BibitemOpen
  \bibfield  {author} {\bibinfo {author} {\bibfnamefont {M.}~\bibnamefont
  {Pieroni}},\ }\href {\doibase 10.1088/1475-7516/2016/02/012} {\bibfield
  {journal} {\bibinfo  {journal} {JCAP}\ }\textbf {\bibinfo {volume} {1602}},\
  \bibinfo {pages} {012} (\bibinfo {year} {2016})},\ \Eprint
  {http://arxiv.org/abs/1510.03691} {arXiv:1510.03691 [hep-ph]} \BibitemShut
  {NoStop}%
\bibitem [{\citenamefont {J{\"a}rv}\ \emph {et~al.}(2017)\citenamefont
  {J{\"a}rv}, \citenamefont {Kannike}, \citenamefont {Marzola}, \citenamefont
  {Racioppi}, \citenamefont {Raidal}, \citenamefont {Rünkla}, \citenamefont
  {Saal},\ and\ \citenamefont {Veerm{\"a}e}}]{Jarv:2016sow}%
  \BibitemOpen
  \bibfield  {author} {\bibinfo {author} {\bibfnamefont {L.}~\bibnamefont
  {J{\"a}rv}}, \bibinfo {author} {\bibfnamefont {K.}~\bibnamefont {Kannike}},
  \bibinfo {author} {\bibfnamefont {L.}~\bibnamefont {Marzola}}, \bibinfo
  {author} {\bibfnamefont {A.}~\bibnamefont {Racioppi}}, \bibinfo {author}
  {\bibfnamefont {M.}~\bibnamefont {Raidal}}, \bibinfo {author} {\bibfnamefont
  {M.}~\bibnamefont {Rünkla}}, \bibinfo {author} {\bibfnamefont
  {M.}~\bibnamefont {Saal}}, \ and\ \bibinfo {author} {\bibfnamefont
  {H.}~\bibnamefont {Veerm{\"a}e}},\ }\href {\doibase
  10.1103/PhysRevLett.118.151302} {\bibfield  {journal} {\bibinfo  {journal}
  {Phys. Rev. Lett.}\ }\textbf {\bibinfo {volume} {118}},\ \bibinfo {pages}
  {151302} (\bibinfo {year} {2017})},\ \Eprint
  {http://arxiv.org/abs/1612.06863} {arXiv:1612.06863 [hep-ph]} \BibitemShut
  {NoStop}%
\bibitem [{\citenamefont {Salvio}(2017)}]{Salvio:2017xul}%
  \BibitemOpen
  \bibfield  {author} {\bibinfo {author} {\bibfnamefont {A.}~\bibnamefont
  {Salvio}},\ }\href {\doibase 10.1140/epjc/s10052-017-4825-6} {\bibfield
  {journal} {\bibinfo  {journal} {Eur. Phys. J.}\ }\textbf {\bibinfo {volume}
  {C77}},\ \bibinfo {pages} {267} (\bibinfo {year} {2017})},\ \Eprint
  {http://arxiv.org/abs/1703.08012} {arXiv:1703.08012 [astro-ph.CO]}
  \BibitemShut {NoStop}%
\bibitem [{\citenamefont {Rasanen}\ and\ \citenamefont
  {Wahlman}(2017)}]{Rasanen:2017ivk}%
  \BibitemOpen
  \bibfield  {author} {\bibinfo {author} {\bibfnamefont {S.}~\bibnamefont
  {Rasanen}}\ and\ \bibinfo {author} {\bibfnamefont {P.}~\bibnamefont
  {Wahlman}},\ }\href {\doibase 10.1088/1475-7516/2017/11/047} {\bibfield
  {journal} {\bibinfo  {journal} {JCAP}\ }\textbf {\bibinfo {volume} {1711}},\
  \bibinfo {pages} {047} (\bibinfo {year} {2017})},\ \Eprint
  {http://arxiv.org/abs/1709.07853} {arXiv:1709.07853 [astro-ph.CO]}
  \BibitemShut {NoStop}%
\bibitem [{\citenamefont {Tenkanen}(2017)}]{Tenkanen:2017jih}%
  \BibitemOpen
  \bibfield  {author} {\bibinfo {author} {\bibfnamefont {T.}~\bibnamefont
  {Tenkanen}},\ }\href {\doibase 10.1088/1475-7516/2017/12/001} {\bibfield
  {journal} {\bibinfo  {journal} {JCAP}\ }\textbf {\bibinfo {volume} {1712}},\
  \bibinfo {pages} {001} (\bibinfo {year} {2017})},\ \Eprint
  {http://arxiv.org/abs/1710.02758} {arXiv:1710.02758 [astro-ph.CO]}
  \BibitemShut {NoStop}%
\bibitem [{\citenamefont {Birrell}\ and\ \citenamefont
  {Davies}(1984)}]{Birrell:1982ix}%
  \BibitemOpen
  \bibfield  {author} {\bibinfo {author} {\bibfnamefont {N.~D.}\ \bibnamefont
  {Birrell}}\ and\ \bibinfo {author} {\bibfnamefont {P.~C.~W.}\ \bibnamefont
  {Davies}},\ }\href {\doibase 10.1017/CBO9780511622632} {\emph {\bibinfo
  {title} {{Quantum Fields in Curved Space}}}},\ Cambridge Monographs on
  Mathematical Physics\ (\bibinfo  {publisher} {Cambridge Univ. Press},\
  \bibinfo {address} {Cambridge, UK},\ \bibinfo {year} {1984})\BibitemShut
  {NoStop}%
\bibitem [{\citenamefont {Tamanini}\ and\ \citenamefont
  {Contaldi}(2011)}]{Tamanini:2010uq}%
  \BibitemOpen
  \bibfield  {author} {\bibinfo {author} {\bibfnamefont {N.}~\bibnamefont
  {Tamanini}}\ and\ \bibinfo {author} {\bibfnamefont {C.~R.}\ \bibnamefont
  {Contaldi}},\ }\href {\doibase 10.1103/PhysRevD.83.044018} {\bibfield
  {journal} {\bibinfo  {journal} {Phys. Rev.}\ }\textbf {\bibinfo {volume}
  {D83}},\ \bibinfo {pages} {044018} (\bibinfo {year} {2011})},\ \Eprint
  {http://arxiv.org/abs/1010.0689} {arXiv:1010.0689 [gr-qc]} \BibitemShut
  {NoStop}%
\bibitem [{\citenamefont {Bauer}\ and\ \citenamefont
  {Demir}(2011)}]{Bauer:2010jg}%
  \BibitemOpen
  \bibfield  {author} {\bibinfo {author} {\bibfnamefont {F.}~\bibnamefont
  {Bauer}}\ and\ \bibinfo {author} {\bibfnamefont {D.~A.}\ \bibnamefont
  {Demir}},\ }\href {\doibase 10.1016/j.physletb.2011.03.042} {\bibfield
  {journal} {\bibinfo  {journal} {Phys. Lett.}\ }\textbf {\bibinfo {volume}
  {B698}},\ \bibinfo {pages} {425} (\bibinfo {year} {2011})},\ \Eprint
  {http://arxiv.org/abs/1012.2900} {arXiv:1012.2900 [hep-ph]} \BibitemShut
  {NoStop}%
\bibitem [{\citenamefont {Markkanen}\ \emph {et~al.}(2017)\citenamefont
  {Markkanen}, \citenamefont {Tenkanen}, \citenamefont {Vaskonen},\ and\
  \citenamefont {Veermäe}}]{Markkanen:2017tun}%
  \BibitemOpen
  \bibfield  {author} {\bibinfo {author} {\bibfnamefont {T.}~\bibnamefont
  {Markkanen}}, \bibinfo {author} {\bibfnamefont {T.}~\bibnamefont {Tenkanen}},
  \bibinfo {author} {\bibfnamefont {V.}~\bibnamefont {Vaskonen}}, \ and\
  \bibinfo {author} {\bibfnamefont {H.}~\bibnamefont {Veermäe}},\ }\href@noop
  {} {\  (\bibinfo {year} {2017})},\ \Eprint {http://arxiv.org/abs/1712.04874}
  {arXiv:1712.04874 [gr-qc]} \BibitemShut {NoStop}%
\bibitem [{\citenamefont {Järv}\ \emph {et~al.}(2017)\citenamefont {Järv},
  \citenamefont {Racioppi},\ and\ \citenamefont {Tenkanen}}]{Jarv:2017azx}%
  \BibitemOpen
  \bibfield  {author} {\bibinfo {author} {\bibfnamefont {L.}~\bibnamefont
  {Järv}}, \bibinfo {author} {\bibfnamefont {A.}~\bibnamefont {Racioppi}}, \
  and\ \bibinfo {author} {\bibfnamefont {T.}~\bibnamefont {Tenkanen}},\
  }\href@noop {} {\  (\bibinfo {year} {2017})},\ \Eprint
  {http://arxiv.org/abs/1712.08471} {arXiv:1712.08471 [gr-qc]} \BibitemShut
  {NoStop}%
\bibitem [{\citenamefont {Kannike}\ \emph {et~al.}(2014)\citenamefont
  {Kannike}, \citenamefont {Racioppi},\ and\ \citenamefont
  {Raidal}}]{Kannike:2014mia}%
  \BibitemOpen
  \bibfield  {author} {\bibinfo {author} {\bibfnamefont {K.}~\bibnamefont
  {Kannike}}, \bibinfo {author} {\bibfnamefont {A.}~\bibnamefont {Racioppi}}, \
  and\ \bibinfo {author} {\bibfnamefont {M.}~\bibnamefont {Raidal}},\ }\href
  {\doibase 10.1007/JHEP06(2014)154} {\bibfield  {journal} {\bibinfo  {journal}
  {JHEP}\ }\textbf {\bibinfo {volume} {06}},\ \bibinfo {pages} {154} (\bibinfo
  {year} {2014})},\ \Eprint {http://arxiv.org/abs/1405.3987} {arXiv:1405.3987
  [hep-ph]} \BibitemShut {NoStop}%
\bibitem [{\citenamefont {Marzola}\ \emph {et~al.}(2016)\citenamefont
  {Marzola}, \citenamefont {Racioppi}, \citenamefont {Raidal}, \citenamefont
  {Urban},\ and\ \citenamefont {Veerm{\"a}e}}]{Marzola:2015xbh}%
  \BibitemOpen
  \bibfield  {author} {\bibinfo {author} {\bibfnamefont {L.}~\bibnamefont
  {Marzola}}, \bibinfo {author} {\bibfnamefont {A.}~\bibnamefont {Racioppi}},
  \bibinfo {author} {\bibfnamefont {M.}~\bibnamefont {Raidal}}, \bibinfo
  {author} {\bibfnamefont {F.~R.}\ \bibnamefont {Urban}}, \ and\ \bibinfo
  {author} {\bibfnamefont {H.}~\bibnamefont {Veerm{\"a}e}},\ }\href {\doibase
  10.1007/JHEP03(2016)190} {\bibfield  {journal} {\bibinfo  {journal} {JHEP}\
  }\textbf {\bibinfo {volume} {03}},\ \bibinfo {pages} {190} (\bibinfo {year}
  {2016})},\ \Eprint {http://arxiv.org/abs/1512.09136} {arXiv:1512.09136
  [hep-ph]} \BibitemShut {NoStop}%
\bibitem [{\citenamefont {Marzola}\ and\ \citenamefont
  {Racioppi}(2016)}]{Marzola:2016xgb}%
  \BibitemOpen
  \bibfield  {author} {\bibinfo {author} {\bibfnamefont {L.}~\bibnamefont
  {Marzola}}\ and\ \bibinfo {author} {\bibfnamefont {A.}~\bibnamefont
  {Racioppi}},\ }\href {\doibase 10.1088/1475-7516/2016/10/010} {\bibfield
  {journal} {\bibinfo  {journal} {JCAP}\ }\textbf {\bibinfo {volume} {1610}},\
  \bibinfo {pages} {010} (\bibinfo {year} {2016})},\ \Eprint
  {http://arxiv.org/abs/1606.06887} {arXiv:1606.06887 [hep-ph]} \BibitemShut
  {NoStop}%
\bibitem [{\citenamefont {Dimopoulos}\ \emph {et~al.}(2017)\citenamefont
  {Dimopoulos}, \citenamefont {Owen},\ and\ \citenamefont
  {Racioppi}}]{Dimopoulos:2017xox}%
  \BibitemOpen
  \bibfield  {author} {\bibinfo {author} {\bibfnamefont {K.}~\bibnamefont
  {Dimopoulos}}, \bibinfo {author} {\bibfnamefont {C.}~\bibnamefont {Owen}}, \
  and\ \bibinfo {author} {\bibfnamefont {A.}~\bibnamefont {Racioppi}},\
  }\href@noop {} {\  (\bibinfo {year} {2017})},\ \Eprint
  {http://arxiv.org/abs/1706.09735} {arXiv:1706.09735 [hep-ph]} \BibitemShut
  {NoStop}%
\bibitem [{\citenamefont {Kannike}\ \emph {et~al.}(2015)\citenamefont
  {Kannike}, \citenamefont {H{\"u}tsi}, \citenamefont {Pizza}, \citenamefont
  {Racioppi}, \citenamefont {Raidal}, \citenamefont {Salvio},\ and\
  \citenamefont {Strumia}}]{Kannike:2015apa}%
  \BibitemOpen
  \bibfield  {author} {\bibinfo {author} {\bibfnamefont {K.}~\bibnamefont
  {Kannike}}, \bibinfo {author} {\bibfnamefont {G.}~\bibnamefont {H{\"u}tsi}},
  \bibinfo {author} {\bibfnamefont {L.}~\bibnamefont {Pizza}}, \bibinfo
  {author} {\bibfnamefont {A.}~\bibnamefont {Racioppi}}, \bibinfo {author}
  {\bibfnamefont {M.}~\bibnamefont {Raidal}}, \bibinfo {author} {\bibfnamefont
  {A.}~\bibnamefont {Salvio}}, \ and\ \bibinfo {author} {\bibfnamefont
  {A.}~\bibnamefont {Strumia}},\ }\href {\doibase 10.1007/JHEP05(2015)065}
  {\bibfield  {journal} {\bibinfo  {journal} {JHEP}\ }\textbf {\bibinfo
  {volume} {05}},\ \bibinfo {pages} {065} (\bibinfo {year} {2015})},\ \Eprint
  {http://arxiv.org/abs/1502.01334} {arXiv:1502.01334 [astro-ph.CO]}
  \BibitemShut {NoStop}%
\bibitem [{\citenamefont {Farzinnia}\ and\ \citenamefont
  {Kouwn}(2016)}]{Farzinnia:2015fka}%
  \BibitemOpen
  \bibfield  {author} {\bibinfo {author} {\bibfnamefont {A.}~\bibnamefont
  {Farzinnia}}\ and\ \bibinfo {author} {\bibfnamefont {S.}~\bibnamefont
  {Kouwn}},\ }\href {\doibase 10.1103/PhysRevD.93.063528} {\bibfield  {journal}
  {\bibinfo  {journal} {Phys. Rev.}\ }\textbf {\bibinfo {volume} {D93}},\
  \bibinfo {pages} {063528} (\bibinfo {year} {2016})},\ \Eprint
  {http://arxiv.org/abs/1512.05890} {arXiv:1512.05890 [hep-ph]} \BibitemShut
  {NoStop}%
\bibitem [{\citenamefont {Kannike}\ \emph {et~al.}(2017)\citenamefont
  {Kannike}, \citenamefont {Racioppi},\ and\ \citenamefont
  {Raidal}}]{Kannike:2016jfs}%
  \BibitemOpen
  \bibfield  {author} {\bibinfo {author} {\bibfnamefont {K.}~\bibnamefont
  {Kannike}}, \bibinfo {author} {\bibfnamefont {A.}~\bibnamefont {Racioppi}}, \
  and\ \bibinfo {author} {\bibfnamefont {M.}~\bibnamefont {Raidal}},\ }\href
  {\doibase 10.1016/j.nuclphysb.2017.02.019} {\bibfield  {journal} {\bibinfo
  {journal} {Nucl. Phys.}\ }\textbf {\bibinfo {volume} {B918}},\ \bibinfo
  {pages} {162} (\bibinfo {year} {2017})},\ \Eprint
  {http://arxiv.org/abs/1605.09378} {arXiv:1605.09378 [hep-ph]} \BibitemShut
  {NoStop}%
\bibitem [{\citenamefont {Koivisto}\ and\ \citenamefont
  {Kurki-Suonio}(2006)}]{Koivisto:2005yc}%
  \BibitemOpen
  \bibfield  {author} {\bibinfo {author} {\bibfnamefont {T.}~\bibnamefont
  {Koivisto}}\ and\ \bibinfo {author} {\bibfnamefont {H.}~\bibnamefont
  {Kurki-Suonio}},\ }\href {\doibase 10.1088/0264-9381/23/7/009} {\bibfield
  {journal} {\bibinfo  {journal} {Class. Quant. Grav.}\ }\textbf {\bibinfo
  {volume} {23}},\ \bibinfo {pages} {2355} (\bibinfo {year} {2006})},\ \Eprint
  {http://arxiv.org/abs/astro-ph/0509422} {arXiv:astro-ph/0509422 [astro-ph]}
  \BibitemShut {NoStop}%
\bibitem [{\citenamefont {Prokopec}\ and\ \citenamefont
  {Weenink}(2013)}]{Prokopec:2013zya}%
  \BibitemOpen
  \bibfield  {author} {\bibinfo {author} {\bibfnamefont {T.}~\bibnamefont
  {Prokopec}}\ and\ \bibinfo {author} {\bibfnamefont {J.}~\bibnamefont
  {Weenink}},\ }\href {\doibase 10.1088/1475-7516/2013/09/027} {\bibfield
  {journal} {\bibinfo  {journal} {JCAP}\ }\textbf {\bibinfo {volume} {1309}},\
  \bibinfo {pages} {027} (\bibinfo {year} {2013})},\ \Eprint
  {http://arxiv.org/abs/1304.6737} {arXiv:1304.6737 [gr-qc]} \BibitemShut
  {NoStop}%
\bibitem [{\citenamefont {J{\"a}rv}\ \emph {et~al.}(2015)\citenamefont
  {J{\"a}rv}, \citenamefont {Kuusk}, \citenamefont {Saal},\ and\ \citenamefont
  {Vilson}}]{Jarv:2014hma}%
  \BibitemOpen
  \bibfield  {author} {\bibinfo {author} {\bibfnamefont {L.}~\bibnamefont
  {J{\"a}rv}}, \bibinfo {author} {\bibfnamefont {P.}~\bibnamefont {Kuusk}},
  \bibinfo {author} {\bibfnamefont {M.}~\bibnamefont {Saal}}, \ and\ \bibinfo
  {author} {\bibfnamefont {O.}~\bibnamefont {Vilson}},\ }\href {\doibase
  10.1103/PhysRevD.91.024041} {\bibfield  {journal} {\bibinfo  {journal} {Phys.
  Rev.}\ }\textbf {\bibinfo {volume} {D91}},\ \bibinfo {pages} {024041}
  (\bibinfo {year} {2015})},\ \Eprint {http://arxiv.org/abs/1411.1947}
  {arXiv:1411.1947 [gr-qc]} \BibitemShut {NoStop}%
\bibitem [{\citenamefont {Kuusk}\ \emph
  {et~al.}(2016{\natexlab{a}})\citenamefont {Kuusk}, \citenamefont {J{\"a}rv},\
  and\ \citenamefont {Vilson}}]{Kuusk:2015dda}%
  \BibitemOpen
  \bibfield  {author} {\bibinfo {author} {\bibfnamefont {P.}~\bibnamefont
  {Kuusk}}, \bibinfo {author} {\bibfnamefont {L.}~\bibnamefont {J{\"a}rv}}, \
  and\ \bibinfo {author} {\bibfnamefont {O.}~\bibnamefont {Vilson}},\ }\href
  {\doibase 10.1142/S0217751X16410037} {\bibfield  {journal} {\bibinfo
  {journal} {Int. J. Mod. Phys.}\ }\textbf {\bibinfo {volume} {A31}},\ \bibinfo
  {pages} {1641003} (\bibinfo {year} {2016}{\natexlab{a}})},\ \Eprint
  {http://arxiv.org/abs/1509.02903} {arXiv:1509.02903 [gr-qc]} \BibitemShut
  {NoStop}%
\bibitem [{\citenamefont {Kuusk}\ \emph
  {et~al.}(2016{\natexlab{b}})\citenamefont {Kuusk}, \citenamefont
  {R{\"u}nkla}, \citenamefont {Saal},\ and\ \citenamefont
  {Vilson}}]{Kuusk:2016rso}%
  \BibitemOpen
  \bibfield  {author} {\bibinfo {author} {\bibfnamefont {P.}~\bibnamefont
  {Kuusk}}, \bibinfo {author} {\bibfnamefont {M.}~\bibnamefont {R{\"u}nkla}},
  \bibinfo {author} {\bibfnamefont {M.}~\bibnamefont {Saal}}, \ and\ \bibinfo
  {author} {\bibfnamefont {O.}~\bibnamefont {Vilson}},\ }\href {\doibase
  10.1088/0264-9381/33/19/195008} {\bibfield  {journal} {\bibinfo  {journal}
  {Class. Quant. Grav.}\ }\textbf {\bibinfo {volume} {33}},\ \bibinfo {pages}
  {195008} (\bibinfo {year} {2016}{\natexlab{b}})},\ \Eprint
  {http://arxiv.org/abs/1605.07033} {arXiv:1605.07033 [gr-qc]} \BibitemShut
  {NoStop}%
\bibitem [{\citenamefont {Flanagan}(2004)}]{Flanagan:2004bz}%
  \BibitemOpen
  \bibfield  {author} {\bibinfo {author} {\bibfnamefont {E.~E.}\ \bibnamefont
  {Flanagan}},\ }\href {\doibase 10.1088/0264-9381/21/15/N02} {\bibfield
  {journal} {\bibinfo  {journal} {Class. Quant. Grav.}\ }\textbf {\bibinfo
  {volume} {21}},\ \bibinfo {pages} {3817} (\bibinfo {year} {2004})},\ \Eprint
  {http://arxiv.org/abs/gr-qc/0403063} {arXiv:gr-qc/0403063 [gr-qc]}
  \BibitemShut {NoStop}%
\bibitem [{\citenamefont {Catena}\ \emph {et~al.}(2007)\citenamefont {Catena},
  \citenamefont {Pietroni},\ and\ \citenamefont {Scarabello}}]{Catena:2006bd}%
  \BibitemOpen
  \bibfield  {author} {\bibinfo {author} {\bibfnamefont {R.}~\bibnamefont
  {Catena}}, \bibinfo {author} {\bibfnamefont {M.}~\bibnamefont {Pietroni}}, \
  and\ \bibinfo {author} {\bibfnamefont {L.}~\bibnamefont {Scarabello}},\
  }\href {\doibase 10.1103/PhysRevD.76.084039} {\bibfield  {journal} {\bibinfo
  {journal} {Phys. Rev.}\ }\textbf {\bibinfo {volume} {D76}},\ \bibinfo {pages}
  {084039} (\bibinfo {year} {2007})},\ \Eprint
  {http://arxiv.org/abs/astro-ph/0604492} {arXiv:astro-ph/0604492 [astro-ph]}
  \BibitemShut {NoStop}%
\bibitem [{\citenamefont {Barvinsky}\ \emph {et~al.}(2008)\citenamefont
  {Barvinsky}, \citenamefont {shchik},\ and\ \citenamefont
  {Starobinsky}}]{Barvinsky:2008ia}%
  \BibitemOpen
  \bibfield  {author} {\bibinfo {author} {\bibfnamefont {A.~O.}\ \bibnamefont
  {Barvinsky}}, \bibinfo {author} {\bibfnamefont {A.~{\relax Yu}.}\
  \bibnamefont {shchik}}, \ and\ \bibinfo {author} {\bibfnamefont {A.~A.}\
  \bibnamefont {Starobinsky}},\ }\href {\doibase 10.1088/1475-7516/2008/11/021}
  {\bibfield  {journal} {\bibinfo  {journal} {JCAP}\ }\textbf {\bibinfo
  {volume} {0811}},\ \bibinfo {pages} {021} (\bibinfo {year} {2008})},\ \Eprint
  {http://arxiv.org/abs/0809.2104} {arXiv:0809.2104 [hep-ph]} \BibitemShut
  {NoStop}%
\bibitem [{\citenamefont {De~Simone}\ \emph {et~al.}(2009)\citenamefont
  {De~Simone}, \citenamefont {Hertzberg},\ and\ \citenamefont
  {Wilczek}}]{DeSimone:2008ei}%
  \BibitemOpen
  \bibfield  {author} {\bibinfo {author} {\bibfnamefont {A.}~\bibnamefont
  {De~Simone}}, \bibinfo {author} {\bibfnamefont {M.~P.}\ \bibnamefont
  {Hertzberg}}, \ and\ \bibinfo {author} {\bibfnamefont {F.}~\bibnamefont
  {Wilczek}},\ }\href {\doibase 10.1016/j.physletb.2009.05.054} {\bibfield
  {journal} {\bibinfo  {journal} {Phys. Lett.}\ }\textbf {\bibinfo {volume}
  {B678}},\ \bibinfo {pages} {1} (\bibinfo {year} {2009})},\ \Eprint
  {http://arxiv.org/abs/0812.4946} {arXiv:0812.4946 [hep-ph]} \BibitemShut
  {NoStop}%
\bibitem [{\citenamefont {Barvinsky}\ \emph {et~al.}(2009)\citenamefont
  {Barvinsky}, \citenamefont {Kamenshchik}, \citenamefont {Kiefer},
  \citenamefont {Starobinsky},\ and\ \citenamefont
  {Steinwachs}}]{Barvinsky:2009fy}%
  \BibitemOpen
  \bibfield  {author} {\bibinfo {author} {\bibfnamefont {A.~O.}\ \bibnamefont
  {Barvinsky}}, \bibinfo {author} {\bibfnamefont {A.~{\relax Yu}.}\
  \bibnamefont {Kamenshchik}}, \bibinfo {author} {\bibfnamefont
  {C.}~\bibnamefont {Kiefer}}, \bibinfo {author} {\bibfnamefont {A.~A.}\
  \bibnamefont {Starobinsky}}, \ and\ \bibinfo {author} {\bibfnamefont
  {C.}~\bibnamefont {Steinwachs}},\ }\href {\doibase
  10.1088/1475-7516/2009/12/003} {\bibfield  {journal} {\bibinfo  {journal}
  {JCAP}\ }\textbf {\bibinfo {volume} {0912}},\ \bibinfo {pages} {003}
  (\bibinfo {year} {2009})},\ \Eprint {http://arxiv.org/abs/0904.1698}
  {arXiv:0904.1698 [hep-ph]} \BibitemShut {NoStop}%
\bibitem [{\citenamefont {Steinwachs}\ and\ \citenamefont
  {Kamenshchik}(2011)}]{Steinwachs:2011zs}%
  \BibitemOpen
  \bibfield  {author} {\bibinfo {author} {\bibfnamefont {C.~F.}\ \bibnamefont
  {Steinwachs}}\ and\ \bibinfo {author} {\bibfnamefont {A.~{\relax Yu}.}\
  \bibnamefont {Kamenshchik}},\ }\href {\doibase 10.1103/PhysRevD.84.024026}
  {\bibfield  {journal} {\bibinfo  {journal} {Phys. Rev.}\ }\textbf {\bibinfo
  {volume} {D84}},\ \bibinfo {pages} {024026} (\bibinfo {year} {2011})},\
  \Eprint {http://arxiv.org/abs/1101.5047} {arXiv:1101.5047 [gr-qc]}
  \BibitemShut {NoStop}%
\bibitem [{\citenamefont {Chiba}\ and\ \citenamefont
  {Yamaguchi}(2013)}]{Chiba:2013mha}%
  \BibitemOpen
  \bibfield  {author} {\bibinfo {author} {\bibfnamefont {T.}~\bibnamefont
  {Chiba}}\ and\ \bibinfo {author} {\bibfnamefont {M.}~\bibnamefont
  {Yamaguchi}},\ }\href {\doibase 10.1088/1475-7516/2013/10/040} {\bibfield
  {journal} {\bibinfo  {journal} {JCAP}\ }\textbf {\bibinfo {volume} {1310}},\
  \bibinfo {pages} {040} (\bibinfo {year} {2013})},\ \Eprint
  {http://arxiv.org/abs/1308.1142} {arXiv:1308.1142 [gr-qc]} \BibitemShut
  {NoStop}%
\bibitem [{\citenamefont {George}\ \emph {et~al.}(2014)\citenamefont {George},
  \citenamefont {Mooij},\ and\ \citenamefont {Postma}}]{George:2013iia}%
  \BibitemOpen
  \bibfield  {author} {\bibinfo {author} {\bibfnamefont {D.~P.}\ \bibnamefont
  {George}}, \bibinfo {author} {\bibfnamefont {S.}~\bibnamefont {Mooij}}, \
  and\ \bibinfo {author} {\bibfnamefont {M.}~\bibnamefont {Postma}},\ }\href
  {\doibase 10.1088/1475-7516/2014/02/024} {\bibfield  {journal} {\bibinfo
  {journal} {JCAP}\ }\textbf {\bibinfo {volume} {1402}},\ \bibinfo {pages}
  {024} (\bibinfo {year} {2014})},\ \Eprint {http://arxiv.org/abs/1310.2157}
  {arXiv:1310.2157 [hep-th]} \BibitemShut {NoStop}%
\bibitem [{\citenamefont {Postma}\ and\ \citenamefont
  {Volponi}(2014)}]{Postma:2014vaa}%
  \BibitemOpen
  \bibfield  {author} {\bibinfo {author} {\bibfnamefont {M.}~\bibnamefont
  {Postma}}\ and\ \bibinfo {author} {\bibfnamefont {M.}~\bibnamefont
  {Volponi}},\ }\href {\doibase 10.1103/PhysRevD.90.103516} {\bibfield
  {journal} {\bibinfo  {journal} {Phys. Rev.}\ }\textbf {\bibinfo {volume}
  {D90}},\ \bibinfo {pages} {103516} (\bibinfo {year} {2014})},\ \Eprint
  {http://arxiv.org/abs/1407.6874} {arXiv:1407.6874 [astro-ph.CO]} \BibitemShut
  {NoStop}%
\bibitem [{\citenamefont {Kamenshchik}\ and\ \citenamefont
  {Steinwachs}(2015)}]{Kamenshchik:2014waa}%
  \BibitemOpen
  \bibfield  {author} {\bibinfo {author} {\bibfnamefont {A.~{\relax Yu}.}\
  \bibnamefont {Kamenshchik}}\ and\ \bibinfo {author} {\bibfnamefont {C.~F.}\
  \bibnamefont {Steinwachs}},\ }\href {\doibase 10.1103/PhysRevD.91.084033}
  {\bibfield  {journal} {\bibinfo  {journal} {Phys. Rev.}\ }\textbf {\bibinfo
  {volume} {D91}},\ \bibinfo {pages} {084033} (\bibinfo {year} {2015})},\
  \Eprint {http://arxiv.org/abs/1408.5769} {arXiv:1408.5769 [gr-qc]}
  \BibitemShut {NoStop}%
\bibitem [{\citenamefont {George}\ \emph {et~al.}(2016)\citenamefont {George},
  \citenamefont {Mooij},\ and\ \citenamefont {Postma}}]{George:2015nza}%
  \BibitemOpen
  \bibfield  {author} {\bibinfo {author} {\bibfnamefont {D.~P.}\ \bibnamefont
  {George}}, \bibinfo {author} {\bibfnamefont {S.}~\bibnamefont {Mooij}}, \
  and\ \bibinfo {author} {\bibfnamefont {M.}~\bibnamefont {Postma}},\ }\href
  {\doibase 10.1088/1475-7516/2016/04/006} {\bibfield  {journal} {\bibinfo
  {journal} {JCAP}\ }\textbf {\bibinfo {volume} {1604}},\ \bibinfo {pages}
  {006} (\bibinfo {year} {2016})},\ \Eprint {http://arxiv.org/abs/1508.04660}
  {arXiv:1508.04660 [hep-th]} \BibitemShut {NoStop}%
\bibitem [{\citenamefont {Miao}\ and\ \citenamefont
  {Woodard}(2015)}]{Miao:2015oba}%
  \BibitemOpen
  \bibfield  {author} {\bibinfo {author} {\bibfnamefont {S.~P.}\ \bibnamefont
  {Miao}}\ and\ \bibinfo {author} {\bibfnamefont {R.~P.}\ \bibnamefont
  {Woodard}},\ }\href {\doibase 10.1088/1475-7516/2015/09/022,
  10.1088/1475-7516/2015/9/022} {\bibfield  {journal} {\bibinfo  {journal}
  {JCAP}\ }\textbf {\bibinfo {volume} {1509}},\ \bibinfo {pages} {022}
  (\bibinfo {year} {2015})},\ \Eprint {http://arxiv.org/abs/1506.07306}
  {arXiv:1506.07306 [astro-ph.CO]} \BibitemShut {NoStop}%
\bibitem [{\citenamefont {Inagaki}\ \emph {et~al.}(2015)\citenamefont
  {Inagaki}, \citenamefont {Nakanishi},\ and\ \citenamefont
  {Odintsov}}]{Inagaki:2015fva}%
  \BibitemOpen
  \bibfield  {author} {\bibinfo {author} {\bibfnamefont {T.}~\bibnamefont
  {Inagaki}}, \bibinfo {author} {\bibfnamefont {R.}~\bibnamefont {Nakanishi}},
  \ and\ \bibinfo {author} {\bibfnamefont {S.~D.}\ \bibnamefont {Odintsov}},\
  }\href {\doibase 10.1016/j.physletb.2015.04.038} {\bibfield  {journal}
  {\bibinfo  {journal} {Phys. Lett.}\ }\textbf {\bibinfo {volume} {B745}},\
  \bibinfo {pages} {105} (\bibinfo {year} {2015})},\ \Eprint
  {http://arxiv.org/abs/1502.06301} {arXiv:1502.06301 [hep-ph]} \BibitemShut
  {NoStop}%
\bibitem [{\citenamefont {Burns}\ \emph {et~al.}(2016)\citenamefont {Burns},
  \citenamefont {Karamitsos},\ and\ \citenamefont {Pilaftsis}}]{Burns:2016ric}%
  \BibitemOpen
  \bibfield  {author} {\bibinfo {author} {\bibfnamefont {D.}~\bibnamefont
  {Burns}}, \bibinfo {author} {\bibfnamefont {S.}~\bibnamefont {Karamitsos}}, \
  and\ \bibinfo {author} {\bibfnamefont {A.}~\bibnamefont {Pilaftsis}},\ }\href
  {\doibase 10.1016/j.nuclphysb.2016.04.036} {\bibfield  {journal} {\bibinfo
  {journal} {Nucl. Phys.}\ }\textbf {\bibinfo {volume} {B907}},\ \bibinfo
  {pages} {785} (\bibinfo {year} {2016})},\ \Eprint
  {http://arxiv.org/abs/1603.03730} {arXiv:1603.03730 [hep-ph]} \BibitemShut
  {NoStop}%
\bibitem [{\citenamefont {Hamada}\ \emph {et~al.}(2017)\citenamefont {Hamada},
  \citenamefont {Kawai}, \citenamefont {Nakanishi},\ and\ \citenamefont
  {Oda}}]{Hamada:2016onh}%
  \BibitemOpen
  \bibfield  {author} {\bibinfo {author} {\bibfnamefont {Y.}~\bibnamefont
  {Hamada}}, \bibinfo {author} {\bibfnamefont {H.}~\bibnamefont {Kawai}},
  \bibinfo {author} {\bibfnamefont {Y.}~\bibnamefont {Nakanishi}}, \ and\
  \bibinfo {author} {\bibfnamefont {K.-y.}\ \bibnamefont {Oda}},\ }\href
  {\doibase 10.1103/PhysRevD.95.103524} {\bibfield  {journal} {\bibinfo
  {journal} {Phys. Rev.}\ }\textbf {\bibinfo {volume} {D95}},\ \bibinfo {pages}
  {103524} (\bibinfo {year} {2017})},\ \Eprint
  {http://arxiv.org/abs/1610.05885} {arXiv:1610.05885 [hep-th]} \BibitemShut
  {NoStop}%
\bibitem [{\citenamefont {Fumagalli}\ and\ \citenamefont
  {Postma}(2016)}]{Fumagalli:2016lls}%
  \BibitemOpen
  \bibfield  {author} {\bibinfo {author} {\bibfnamefont {J.}~\bibnamefont
  {Fumagalli}}\ and\ \bibinfo {author} {\bibfnamefont {M.}~\bibnamefont
  {Postma}},\ }\href {\doibase 10.1007/JHEP05(2016)049} {\bibfield  {journal}
  {\bibinfo  {journal} {JHEP}\ }\textbf {\bibinfo {volume} {05}},\ \bibinfo
  {pages} {049} (\bibinfo {year} {2016})},\ \Eprint
  {http://arxiv.org/abs/1602.07234} {arXiv:1602.07234 [hep-ph]} \BibitemShut
  {NoStop}%
\bibitem [{\citenamefont {Fumagalli}(2017)}]{Fumagalli:2016sof}%
  \BibitemOpen
  \bibfield  {author} {\bibinfo {author} {\bibfnamefont {J.}~\bibnamefont
  {Fumagalli}},\ }\href {\doibase 10.1016/j.physletb.2017.04.017} {\bibfield
  {journal} {\bibinfo  {journal} {Phys. Lett.}\ }\textbf {\bibinfo {volume}
  {B769}},\ \bibinfo {pages} {451} (\bibinfo {year} {2017})},\ \Eprint
  {http://arxiv.org/abs/1611.04997} {arXiv:1611.04997 [hep-th]} \BibitemShut
  {NoStop}%
\bibitem [{\citenamefont {Bezrukov}\ \emph {et~al.}(2017)\citenamefont
  {Bezrukov}, \citenamefont {Pauly},\ and\ \citenamefont
  {Rubio}}]{Bezrukov:2017dyv}%
  \BibitemOpen
  \bibfield  {author} {\bibinfo {author} {\bibfnamefont {F.}~\bibnamefont
  {Bezrukov}}, \bibinfo {author} {\bibfnamefont {M.}~\bibnamefont {Pauly}}, \
  and\ \bibinfo {author} {\bibfnamefont {J.}~\bibnamefont {Rubio}},\
  }\href@noop {} {\  (\bibinfo {year} {2017})},\ \Eprint
  {http://arxiv.org/abs/1706.05007} {arXiv:1706.05007 [hep-ph]} \BibitemShut
  {NoStop}%
\bibitem [{\citenamefont {Karam}\ \emph
  {et~al.}(2017{\natexlab{b}})\citenamefont {Karam}, \citenamefont {Pappas},\
  and\ \citenamefont {Tamvakis}}]{Karam:2017zno}%
  \BibitemOpen
  \bibfield  {author} {\bibinfo {author} {\bibfnamefont {A.}~\bibnamefont
  {Karam}}, \bibinfo {author} {\bibfnamefont {T.}~\bibnamefont {Pappas}}, \
  and\ \bibinfo {author} {\bibfnamefont {K.}~\bibnamefont {Tamvakis}},\ }\href
  {\doibase 10.1103/PhysRevD.96.064036} {\bibfield  {journal} {\bibinfo
  {journal} {Phys. Rev.}\ }\textbf {\bibinfo {volume} {D96}},\ \bibinfo {pages}
  {064036} (\bibinfo {year} {2017}{\natexlab{b}})},\ \Eprint
  {http://arxiv.org/abs/1707.00984} {arXiv:1707.00984 [gr-qc]} \BibitemShut
  {NoStop}%
\bibitem [{\citenamefont {Narain}(2017)}]{Narain:2017mtu}%
  \BibitemOpen
  \bibfield  {author} {\bibinfo {author} {\bibfnamefont {G.}~\bibnamefont
  {Narain}},\ }\href {\doibase 10.1088/1475-7516/2017/10/032} {\bibfield
  {journal} {\bibinfo  {journal} {JCAP}\ }\textbf {\bibinfo {volume} {1710}},\
  \bibinfo {pages} {032} (\bibinfo {year} {2017})},\ \Eprint
  {http://arxiv.org/abs/1708.00830} {arXiv:1708.00830 [gr-qc]} \BibitemShut
  {NoStop}%
\bibitem [{\citenamefont {Ruf}\ and\ \citenamefont
  {Steinwachs}(2017)}]{Ruf:2017xon}%
  \BibitemOpen
  \bibfield  {author} {\bibinfo {author} {\bibfnamefont {M.~S.}\ \bibnamefont
  {Ruf}}\ and\ \bibinfo {author} {\bibfnamefont {C.~F.}\ \bibnamefont
  {Steinwachs}},\ }\href@noop {} {\  (\bibinfo {year} {2017})},\ \Eprint
  {http://arxiv.org/abs/1711.07486} {arXiv:1711.07486 [gr-qc]} \BibitemShut
  {NoStop}%
\bibitem [{\citenamefont {Ferreira}\ \emph {et~al.}(2018)\citenamefont
  {Ferreira}, \citenamefont {Hill},\ and\ \citenamefont
  {Ross}}]{Ferreira:2018itt}%
  \BibitemOpen
  \bibfield  {author} {\bibinfo {author} {\bibfnamefont {P.~G.}\ \bibnamefont
  {Ferreira}}, \bibinfo {author} {\bibfnamefont {C.~T.}\ \bibnamefont {Hill}},
  \ and\ \bibinfo {author} {\bibfnamefont {G.~G.}\ \bibnamefont {Ross}},\
  }\href@noop {} {\  (\bibinfo {year} {2018})},\ \Eprint
  {http://arxiv.org/abs/1801.07676} {arXiv:1801.07676 [hep-th]} \BibitemShut
  {NoStop}%
\bibitem [{\citenamefont {Zee}(1979)}]{PhysRevLett.42.417}%
  \BibitemOpen
  \bibfield  {author} {\bibinfo {author} {\bibfnamefont {A.}~\bibnamefont
  {Zee}},\ }\href {\doibase 10.1103/PhysRevLett.42.417} {\bibfield  {journal}
  {\bibinfo  {journal} {Phys. Rev. Lett.}\ }\textbf {\bibinfo {volume} {42}},\
  \bibinfo {pages} {417} (\bibinfo {year} {1979})}\BibitemShut {NoStop}%
\bibitem [{\citenamefont {Smolin}(1979)}]{SMOLIN1979253}%
  \BibitemOpen
  \bibfield  {author} {\bibinfo {author} {\bibfnamefont {L.}~\bibnamefont
  {Smolin}},\ }\href {\doibase https://doi.org/10.1016/0550-3213(79)90059-2}
  {\bibfield  {journal} {\bibinfo  {journal} {Nuclear Physics B}\ }\textbf
  {\bibinfo {volume} {160}},\ \bibinfo {pages} {253 } (\bibinfo {year}
  {1979})}\BibitemShut {NoStop}%
\bibitem [{\citenamefont {Cooper}\ and\ \citenamefont
  {Venturi}(1981)}]{Venturi:1981}%
  \BibitemOpen
  \bibfield  {author} {\bibinfo {author} {\bibfnamefont {F.}~\bibnamefont
  {Cooper}}\ and\ \bibinfo {author} {\bibfnamefont {G.}~\bibnamefont
  {Venturi}},\ }\href {\doibase 10.1103/PhysRevD.24.3338} {\bibfield  {journal}
  {\bibinfo  {journal} {Phys. Rev. D}\ }\textbf {\bibinfo {volume} {24}},\
  \bibinfo {pages} {3338} (\bibinfo {year} {1981})}\BibitemShut {NoStop}%
\bibitem [{\citenamefont {Spokoiny}(1984)}]{SPOKOINY198439}%
  \BibitemOpen
  \bibfield  {author} {\bibinfo {author} {\bibfnamefont {B.}~\bibnamefont
  {Spokoiny}},\ }\href {\doibase https://doi.org/10.1016/0370-2693(84)90587-2}
  {\bibfield  {journal} {\bibinfo  {journal} {Physics Letters B}\ }\textbf
  {\bibinfo {volume} {147}},\ \bibinfo {pages} {39 } (\bibinfo {year}
  {1984})}\BibitemShut {NoStop}%
\bibitem [{\citenamefont {Cervantes-Cota}\ and\ \citenamefont
  {Dehnen}(1995{\natexlab{a}})}]{CervantesCota:1994zf}%
  \BibitemOpen
  \bibfield  {author} {\bibinfo {author} {\bibfnamefont {J.~L.}\ \bibnamefont
  {Cervantes-Cota}}\ and\ \bibinfo {author} {\bibfnamefont {H.}~\bibnamefont
  {Dehnen}},\ }\href {\doibase 10.1103/PhysRevD.51.395} {\bibfield  {journal}
  {\bibinfo  {journal} {Phys. Rev.}\ }\textbf {\bibinfo {volume} {D51}},\
  \bibinfo {pages} {395} (\bibinfo {year} {1995}{\natexlab{a}})},\ \Eprint
  {http://arxiv.org/abs/astro-ph/9412032} {arXiv:astro-ph/9412032 [astro-ph]}
  \BibitemShut {NoStop}%
\bibitem [{\citenamefont {Cervantes-Cota}\ and\ \citenamefont
  {Dehnen}(1995{\natexlab{b}})}]{CervantesCota:1995tz}%
  \BibitemOpen
  \bibfield  {author} {\bibinfo {author} {\bibfnamefont {J.~L.}\ \bibnamefont
  {Cervantes-Cota}}\ and\ \bibinfo {author} {\bibfnamefont {H.}~\bibnamefont
  {Dehnen}},\ }\href {\doibase 10.1016/0550-3213(95)00128-X} {\bibfield
  {journal} {\bibinfo  {journal} {Nucl. Phys.}\ }\textbf {\bibinfo {volume}
  {B442}},\ \bibinfo {pages} {391} (\bibinfo {year} {1995}{\natexlab{b}})},\
  \Eprint {http://arxiv.org/abs/astro-ph/9505069} {arXiv:astro-ph/9505069
  [astro-ph]} \BibitemShut {NoStop}%
\bibitem [{\citenamefont {Lyth}\ and\ \citenamefont
  {Riotto}(1999)}]{Lyth:1998xn}%
  \BibitemOpen
  \bibfield  {author} {\bibinfo {author} {\bibfnamefont {D.~H.}\ \bibnamefont
  {Lyth}}\ and\ \bibinfo {author} {\bibfnamefont {A.}~\bibnamefont {Riotto}},\
  }\href {\doibase 10.1016/S0370-1573(98)00128-8} {\bibfield  {journal}
  {\bibinfo  {journal} {Phys. Rept.}\ }\textbf {\bibinfo {volume} {314}},\
  \bibinfo {pages} {1} (\bibinfo {year} {1999})},\ \Eprint
  {http://arxiv.org/abs/hep-ph/9807278} {arXiv:hep-ph/9807278 [hep-ph]}
  \BibitemShut {NoStop}%
\end{thebibliography}%

\end{document}